\documentclass{aa}
\usepackage[varg]{txfonts}
\usepackage{natbib}
\usepackage[english]{babel}
\usepackage[utf8]{inputenc}
\usepackage{soul}
\usepackage{graphicx}
\usepackage{txfonts}
\usepackage{color}
\usepackage[dvipsnames]{xcolor}

\begin{document}

\title{Unveiling pure-metal ejecta X-ray emission in supernova remnants through their radiative recombination continuum}
\titlerunning{RRC in Pure-metal}
\author{Emanuele Greco\inst{1,2,3} \and Jacco Vink\inst{3,4,5} \and Marco Miceli\inst{1,2} \and Salvatore Orlando\inst{2} \and Vladimir Dom{\v c}ek\inst{3,4} \and Ping Zhou\inst{3}  \and Fabrizio Bocchino\inst{2} \and Giovanni Peres\inst{1,2}}
\authorrunning{Greco et al.}
\institute{Dipartimento di Fisica e Chimica, Università degli Studi di Palermo, Piazza del Parlamento 1, 90134, Palermo, Italy \and INAF-Osservatorio Astronomico di Palermo, Piazza del Parlamento 1, 90134, Palermo, Italy \and Anton Pannekoek Institute for Astronomy, University of Amsterdam, Science Park 904, 1098 XH Amsterdam, The Netherlands \and GRAPPA, University of Amsterdam, Science Park 904, 1098 XH Amsterdam, The Netherlands \and SRON, Netherlands Institute for Space Research, Utrech, The Netherlands}

\date{Last compile: \today}
\abstract{\\ \emph{Context}. Spectral analysis of X-ray emission from ejecta in supernova remnants (SNRs) is hampered by the low spectral resolution of CCD detectors, which typically creates a degeneracy between the best-fit values of chemical abundances and the plasma emission measure. The combined contribution of shocked ambient medium and ejecta to the emerging X-ray emission further complicates the determination of the ejecta mass and chemical composition. This degeneracy leads to big uncertainties in mass estimates and can introduce a bias in the comparison between the ejecta chemical composition derived from the observations and the yields predicted by explosive nucleosynthesis models.  \\ \emph{Aims.} We explore the capabilities of present and future spectral instruments with the aim of identifying a spectral feature which may allow us to discriminate between metal-rich and pure-metal plasmas in X-ray spectra of SNRs.\\ \emph{Methods.}
We studied the behavior of the most common X-ray emission processes of an optically thin plasma in the high-abundance regime. We investigated spectral features of bremsstrahlung, radiative recombination continua (RRC) and line emission, by exploring a wide range of chemical abundances, plasma temperatures and ionization parameters. We then synthesized X-ray spectra from a state-of-the-art 3D hydrodynamic (HD) simulation of Cas A, by using the response matrix from the \emph{Chandra} ACIS-S CCD detector and that of the XRISM/Resolve X-ray calorimeter spectrometer.\\ \emph{Results.} We found that a bright RRC shows up when the plasma is made of pure-metal ejecta, and a high spectral resolution is needed to actually identify this ejecta signature. We tested and verified the applicability of our novel diagnostic tool and we propose a promising target for the future detection of such spectral feature: the southeastern Fe-rich clump of Cas A.\\ \emph{Conclusions.} While there is no way to unambiguously reveal pure-metal ejecta emission with CCD detectors, X-ray calorimeters will be able to pinpoint the presence of pure-metal RRC and to recover correctly absolute mass and the chemical composition of the ejecta, opening a new window on the link progenitor-supernova-remnant.\\

}

\maketitle             

\section{Introduction}
\label{intro}
Supernova Remnants (SNRs), the outcome of supernova (SN) explosions, are extended sources with a complex morphology which depends on the properties of the progenitor star and of the ambient medium, and on the processes associated with the SN explosion. After the explosion, the ejecta of the progenitor star expand behind the SN blast wave which interacts with the Interstellar Medium (ISM) by heating it up to X-ray emitting temperatures. As the forward shock moves outwards, a reverse shock penetrates and heats the cold expanding ejecta to X-ray emitting temperature, leading to additional X-ray emission. 

X-ray spectral analysis of SNRs is a powerful tool to study chemical composition and mass of the shocked ejecta. Having information on these quantities is fundamental to know more about the progenitor star, the SN explosion, and the explosive nucleosynthesis processes. X-ray spectra  of SNRs typically also include the contribution of the shocked ISM, thus making it more difficult to disentangle the contributions of the continuum and line emission from the ejecta and the swept-up gas, and to infer their physical and chemical properties. Moreover, X-ray spectral analysis of SNRs performed with CCD detectors with moderate spectral resolution ($R\sim 5-100$) is usually affected by a degeneracy between the best-fit values of chemical abundances and the plasma emission measure.  In fact, because of the low energy resolution, the blending between different emission lines can create a ``false continuum", that makes it difficult to constrain the real continuum flux, especially below 4 keV. Thus, it is possible to describe a certain spectrum either with high abundances and low emission measure or vice versa. This degeneracy leads to big uncertainties in the mass estimates and may even hide the existence of pure-metal ejecta plasma in SNRs (\citealt{vkb96}).

Such uncertainty can be partially addressed by measuring relative abundances between elements, typically by adopting Si or Fe as a reference (e.g. \citealt{wbv02}, \citealt{mdb06,mbr08}, \citealt{ksg12}, \citealt{lcs14}, \citealt{fbp15}, \citealt{zv18}, \citealt{zvs19}). On the other hand, this approach does not allow us to unambiguously derive absolute mass estimates for the yields in SNe. The comparison with theoretical nucleosynthesis yields (e.g. \citealt{ww95}, \citealt{tnh96}, \citealt{nht97}, \citealt{num99}, \citealt{sew16}) can be only performed through abundances ratios and may lead to a misunderstanding of the effective explosion mechanism or of the actual progenitor star properties. In order to have a fully reliable estimate of the abundance and of the mass of each element and to correctly compare these values with the theoretical predictions, a tool able to precisely estimate the absolute abundances of ejecta is required.

In this paper, we present a study on the behaviour in the high-abundance regime of the main emission processes in SNRs: Bremmstrahlung (Free-Free process, indicated with FF), Radiative Recombination Continua (Free-Bound process, RRC or FB) and Line emission. We performed a set of spectral simulations in order to identify a signature of pure-metal ejecta emission in the X-ray spectra of SNRs. We identified a novel diagnostic tool and successfully tested its applicability by adopting the Galactic SNR Cassiopeia A (hereafter Cas A) as a benchmark. 

The paper is organized as follows: we identify and describe the nature of the pure-metal ejecta signature in Sect. \ref{ss}; in Sect. \ref{allsynthesis} we describe the synthesis method and the resulting synthetic spectra; in Sect. \ref{introCasA} we investigate the possible detection of pure-metal ejecta emission in Cas A; in Sect. \ref{final} we discuss our results and summarize our conclusions.

\section{X-ray emission of metal-rich plasma}
\label{ss}

\subsection{Main X-ray emission processes}
The X-ray emitting plasma in SNRs is optically thin and its emission depends on binary collisions between electrons and ions. At a fixed temperature, $T$, the emissivity is determined by the Emission Measure (EM), defined as $EM=\int{n_e n_H dV}$ where $n_e$ is the electron density and $n_H$ is the ion density of hydrogen (H) (for solar abundances, $n_e \approx n_H$). In the following, we briefly recall the emissivity equations for all the emission processes mentioned in Sect. \ref{intro}.

\label{line}
The emissivity of the line emission for a particular transition $j\rightarrow i$ of an element $z$ is (\citealt{mew99}):
\begin{equation}
    P_{ji} \propto A_z n_H n_e F(T) \,\,\, \mathrm{photons \cdot cm^{-3}s^{-1}}
    \label{eqline}
\end{equation}
where $A_z$ is the abundance of the element $z$ and $F(T)$ is a function, indicating the temperature dependence due to the combined effects of ionization and excitation, which sharply peaks at the characteristic line temperature.

\label{bs}
The total bremsstrahlung emissivity is given by the sum of the emissivities of all ion species (Eq. \ref{eqff}, \citealt{mew99}):

\begin{equation}
    \varepsilon_{ff} \propto \sum_i Z_i^2 n_e n_i T^{-1/2} \exp\left(\frac{-h\nu}{kT}\right) \,\,\, \mathrm{erg \, cm ^{-3} s^{-1} Hz^{-1}}
    \label{eqff}
\end{equation}
where $Z_i$ and $n_i$ are the effective charge and density of the ion $i$.
In a plasma with solar (or mildly enhanced) abundances, the main contribution to the bremsstrahlung emission originates from H ions and electrons stripped from H atoms, since this element is by far the most abundant one (values of the proto-solar abundances in logarithmic units according to \citealt{lpg09} are shown in Table \ref{tabab}). In the pure-metal ejecta (or extremely metal-rich) scenario, instead, we expect the heavy ions and electrons stemming from them to significantly contribute to the FF emission.

\begin{table}[!ht]
\centering
\begin{tabular}{c|c}
\hline\hline
Element & Abundance (log units)\\
\hline
H& 12 \\
\hline
He& 10.987\\
\hline
C&8.443\\
\hline
O&8.782\\
\hline
Ne& 8.103\\
\hline
Mg& 7.599\\
\hline
Si& 7.586\\
\hline
S& 7.210\\
\hline 
Ar& 6.553\\
\hline 
Ca& 6.367\\
\hline 
Fe& 7.514\\
\hline
Ni& 6.276\\
\end{tabular}
\caption{Proto-solar abundances, expressed in logarithmic
units with H=12.0, by definition, according to \cite{lpg09}.}
\label{tabab}
\end{table}
\label{fb}
 Eq. \ref{eqfb} shows the emissivity for the FB process (\citealt{lie1999}):

\begin{equation}
    \varepsilon_{fb} \propto n_e n_{i+1} \left(\frac{h\nu}{kT^{1/2}}\right)^3 \exp\left({\frac{-h\nu+\chi}{kT}}\right)  \,\,\, \mathrm{erg \, cm^{-3} s^{-1} Hz^{-1}}
    \label{eqfb}
\end{equation}

\noindent where $n_{i+1}$ is the ion density of the recombining ion, $h\nu$ in the X-ray photon energy, and $\chi$ is the ionization energy. The width of the RRC is $\Delta\nu\approx kT$ and if $ kT << h\nu $ the emission results in relatively narrow, line-like, emission peaks near the series limits of lines. If $kT>> h\nu$ the RRC is wide and looks like a continuum distribution, similar in shape to FF above the RRC edge. Up to now, this contribution has been observed only in recombining (overionized) plasmas like those detected in a large sample of mixed morphology SNRs during the last decade (e.g. \citealt{yok09}, \citealt{mtu17}, \citealt{gmo18} for IC443; \citealt{oky09}, \citealt{mbd10} for W49b, \citealt{zmb11}; \citealt{uky12} for W44; \citealt{sk12} for W28). 

\subsection{X-ray spectral signatures of pure-metal ejecta}
\label{ss2}
We performed spectral simulations using the X-ray spectral analysis code SPEX (version 3.04.00 with SPEXACT 2.07.00, \citealt{spex}) with the aim of investigating the behavior of the X-ray emission processes at high abundances.
Considering a plasma in collisional ionization equilibrium (CIE), we studied the dependence of the flux on the abundance of two elements: silicon (Si) and iron (Fe). For our simulations, we considered the energy ranges reported in Table \ref{tab:bands}\footnote{The S abundance, as well the abundance of all other elements, is fixed to 1. Thus, even if the S XVI line is present at 2.623 keV, it does not affect our results.}.
\begin{table}[!ht]
\centering
\begin{tabular}{c|c}
\hline\hline
Band name & Energy range (keV)\\
\hline
SiLine & 1.79-2.01 \\
SiCont & 2.47-2.67 \\ 
\hline
FeLine & 1.2-1.4 \\
FeCont & 2.05-2.15 \\
\hline
\end{tabular}
\caption{Energy bands adopted for the X-ray fluxes in Fig. \ref{Flux_comparison}.}
\label{tab:bands}
\end{table}
The SiLine band includes the Si XIII (1.865 keV) and Si XIV (2.006 keV) emission lines, while the SiCont band was chosen to cover the emission at energies slightly above that of the Si XIII recombination edge (2.438 keV) and excluding the Si XIV recombination edge (2.673 keV). FeLine band includes the forest of Fe L-lines for ions Fe XVII-XXIII, while FeCont band covers the emission at energies above the Fe XXIV recombination edge (2.023 keV). Values of the ionization energy are taken from \cite{lid03} and the lines energies are taken from AtomDB WebGuide\footnote{http://www.atomdb.org/Webguide/webguide.php.}.

In the following, we present the results obtained for Si for the sake of clarity, since the Fe emission shows a very complex line pattern up to energies close to those of its RRC. The results obtained for Fe are analogous to those obtained for Si and are described in detail in Appendix A.

Figure \ref{Flux_comparison} shows the integrated flux for each emission process in the corresponding energy band, normalized to that obtained for solar abundances, for a plasma temperature of 1 keV. The flux of line emission increases linearly with the abundance (as predicted by Eq. \ref{eqline}). The total FB emission shows a weak increase for abundances values between 1 and 10 because the FB emission associated with Si is only a fraction of the total FB emission. Once $ A_{Si} \ga $ 10, the total FB emission is due mainly to the Si FB and we observe a linear dependence (as predicted by Eq. \ref{eqfb}). The FF emission is substantially insensitive to the increasing abundance until values of a few hundreds are reached; then, it increases with the abundance as the FB and line processes.
 This is because, for abundances $A_{Si}\la 10^2$, the FF emission produced by electrons originally belonging to H (hereafter H-electrons), overcomes that of electrons stripped from Si (Si-electrons). For solar abundances the number of H-electrons is about $10^4$ times the number of Si-electrons (Table \ref{tabab} and Eq. \ref{eqff}), therefore, by only slightly increasing the Si abundance, the global contribution to the FF emission is still mainly associated with H-electrons (and H ions). 

As abundances of the order of a few hundreds are reached, the number of Si-electrons is not negligible with respect to the number of H-electrons. Moreover, in this regime, that we call \emph{pure-metal ejecta} regime, the contribution of Si ions to the electron scattering becomes important. For these high abundances, therefore, the term including the Si contribution becomes the dominant one in the summation of  Eq. \ref{eqff} (given the dependance on $Z_i^2$, with Z=12 for  He-like Si) and we thus observe the expected linear increase with the abundance. Figure \ref{RatioFBFF} shows the FB over FF flux ratio: the observed flattening in the flux ratio reflects the pure-metal ejecta regime discussed so far.

 We also investigated how the FB to FF flux ratio depends on the plasma  temperature. We repeated the simulations described above by exploring temperatures in the range $kT=0.2-3$ keV with a step of 0.1 keV. Figure \ref{RatioFBFF2} shows the FB to FF flux ratios for three different temperatures, namely $kT=0.2$ keV, $kT=0.9$ keV and $kT=1.7$ keV. By increasing the temperature in the range 0.2 to 0.8 keV, the slope $\sigma$ of the FB to FF ratio as a function of Si abundance increases; i.e., with high plasma temperature, the FB contribution becomes more visible over the FF. However, at $kT\sim$ 0.9 keV, $\sigma$ reaches its maximum and a further increase in temperature leads to lower values of $\sigma$. The observed trend at energies below 0.9 keV is due to the increasing degree of Si ionization and the subsequent higher number of free electrons combined with the increasing width of the RRC, which is spread out at higher temperatures reducing the effect in the 2.47-2.67 keV band. On the other hand, at energies above this threshold, the electrons are so energetic that they can escape recombination while still increasing the FF emission. The value of the temperature threshold depends on the element considered, because of the different degree of ionization at a given temperature and of the corresponding number of vacancies in the ion itself. Heavier elements have higher thresholds (see Fig. \ref{fluxfe} for the Fe case) also because of the stronger ion electrostatic field. We found that, in the Si case, the FB to FF ratio is maximized when the electron temperature is in the range $kT=0.6-1.2$ keV. In the Fe case, the corresponding range is $kT=1.5-3.5$ keV (see Appendix A).
 
In conclusion, we expect that the spectral signature of pure-metal ejecta must be related to the enhanced FB emission. Therefore, a careful study of  edges continuum can be adopted as a diagnostic tool to reveal pure metal ejecta in SNRs.

\begin{figure}[!ht]
    \centering
    \includegraphics[scale=0.15]{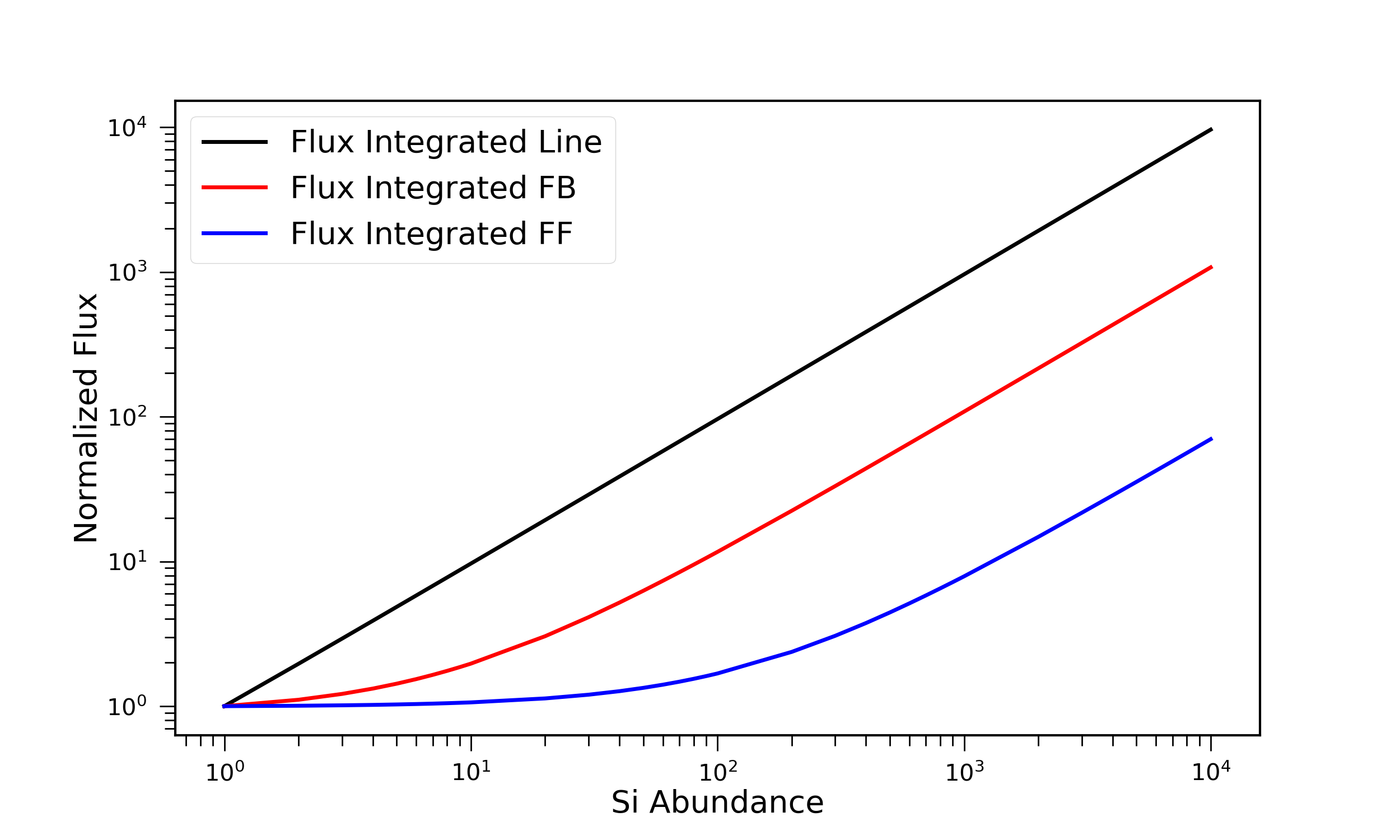}
    \caption{X-ray fluxes of different emission processes (in the corresponding energy bands, see Tab. \ref{tab:bands}) as a function of Si abundance for a plasma temperature of 1 keV. All values are normalized to those obtained for solar abundances.} 
    \label{Flux_comparison}
\end{figure}

\begin{figure}[!ht]
    \centering
    \includegraphics[scale=0.15]{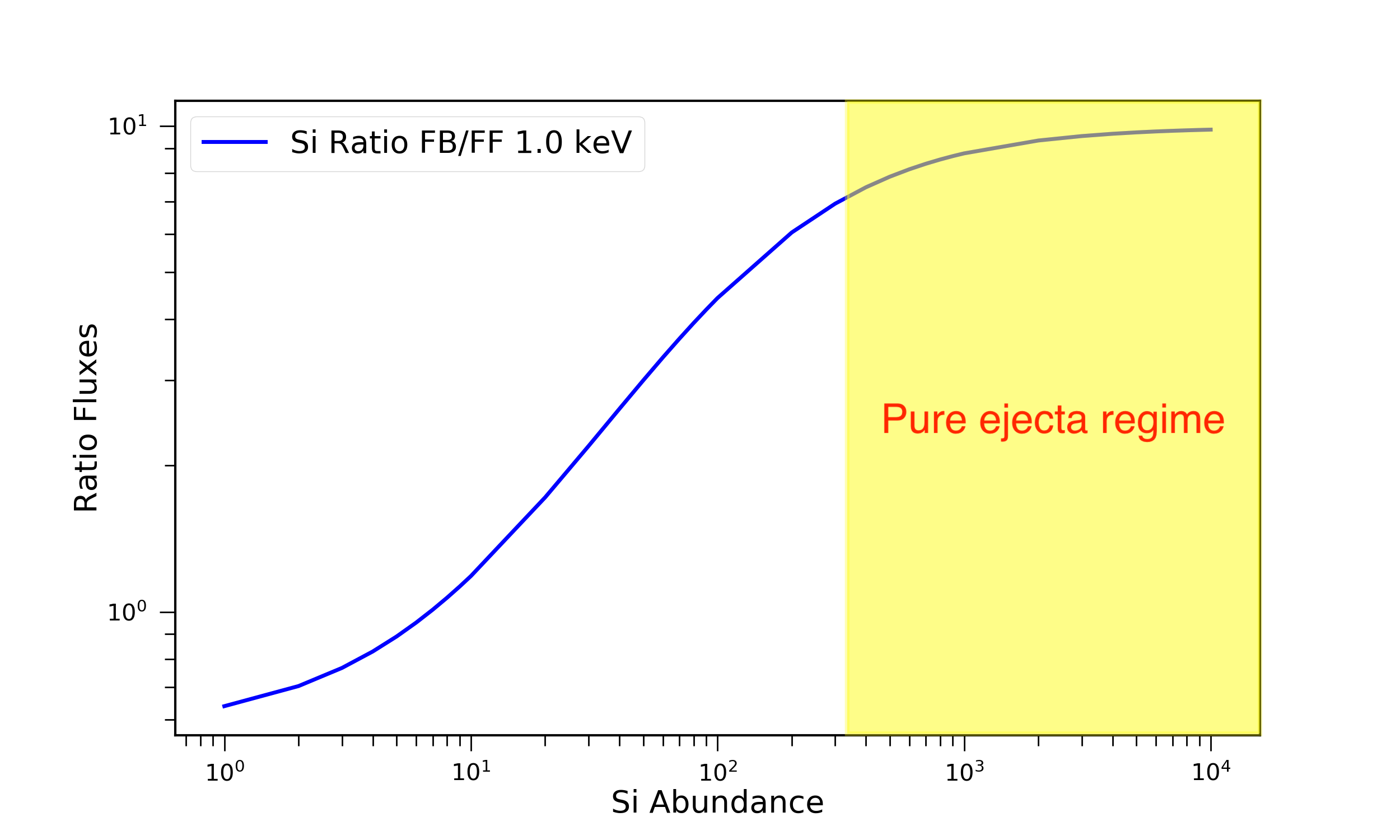}
    \caption{Ratio between the FF and FB fluxes shown in Fig. \ref{Flux_comparison} as a function of the Si abundance for a plasma temperature of 1 keV.}
    \label{RatioFBFF}
\end{figure}

\begin{figure}[!ht]
    \centering

    \includegraphics[scale=0.15]{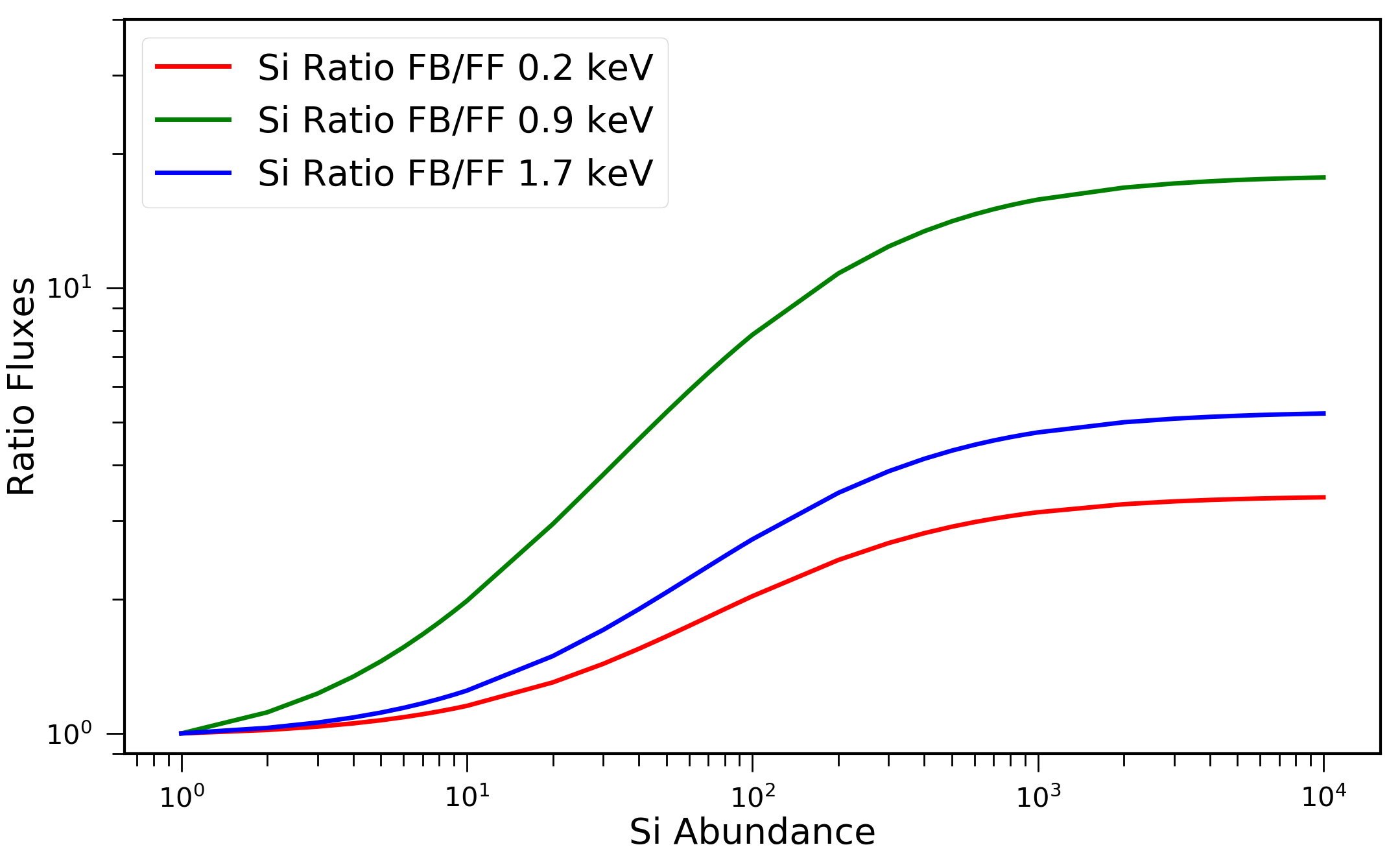}
    \caption{Same as Fig. \ref{RatioFBFF} for plasma temperatures of 0.2 keV (red line), 0.9 keV (green) and 1.7 keV (blue).}
    \label{RatioFBFF2}
\end{figure}

\section{Synthetic X-ray spectra}
\label{allsynthesis}

To investigate the observability of pure-metal ejecta emission in SNRs, we produced synthetic spectra by folding our spectral models with the response matrix of actual detectors.
We also included the contribution of the ISM X-ray emission. Pure-metal ejecta can be distributed on large, expanding shells as well as concentrated in dense clumps embedded in an environment of shocked ISM, where the mixing between ISM and ejecta is less remarkable (as \citealt{hl03} found for the Fe cloudlets in Cas A). In any case, the ejecta emission is always superimposed to the emission stemming from the shocked ISM. Through this whole section, we assume a generical value for the column density equal to $n_H=5\times10^{21}$ cm$^{-2}$, unless otherwise stated.

For our simulations, we used the response matrices of ACIS-S \footnote{Using the XMM-Newton/MOS response matrix instead of the CHANDRA/ACIS-S leads to identical results.}, a CCD detector on board the \emph{Chandra} X-ray telescope\footnote{https://heasarc.gsfc.nasa.gov/docs/chandra/chandra.html} and Resolve, a microcalorimeter that will be onboard
\emph{XRISM}\footnote{http://xrism.isas.jaxa.jp}, the JAXA/NASA X-ray telescope expected to be launched in 2021. The main instrumental characteristics of these intruments are summarized in Table \ref{comparisonresponse}. \emph{XRISM} will have lower effective area and spatial resolution than \emph{Chandra} but, thanks to the microcalorimeter, its spectral resolution will be better than that of \emph{Chandra} by a factor of 10. 

\begin{table}[!ht]
\centering
\begin{tabular}{c|c|c}
\hline\hline
Detector & ACIS-S & Resolve \\
Technology  & CCD & Microcal. \\
\hline
Field of view & 9'x9'& 3'x3'  \\
PSF & 0.5''& 1.7' \\
Energy res. @ 6 keV (eV) & 50 & 5  \\
Eff. Area @ 1 keV (cm$^2$) & 340& 160 \\
Eff. Area @ 6 keV (cm$^2$) & 230& 210  \\
Energy range (keV) & 0.15-12 & 0.3-12 \\
\hline
\end{tabular}
\caption{Main characteristics of ACIS-S (\emph{Chandra}) and Resolve (XRISM) detectors.}
\label{comparisonresponse}
\end{table}

\subsection{ACIS-S}
\label{chandrasynth}
\begin{figure}[!ht]
    \centering
    \includegraphics[scale=0.24]{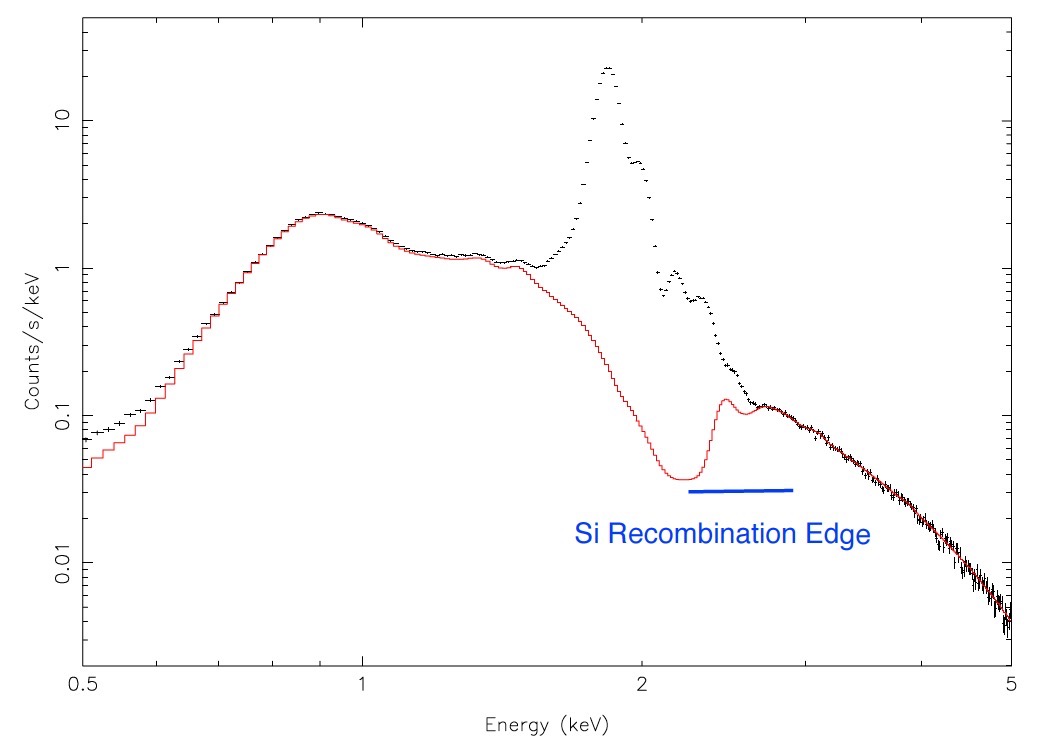}
    \caption{Black: synthetic ACIS-S spectrum of a CIE plasma with abundance of all elements set to 3, with the exception of Si abundance set to 300, $kT=0.8$ keV and EM=1.5$\times10^{55}$ cm$^{-3}$. Red solid: same spectrum but with the Si line emission subtracted but Si contributions to the FF and FB continuum are still included.}
    \label{lineonly}
\end{figure}

Here, we show that CCD detectors are not able to detect the RRC edge. To do that, we simulated a \emph{Chandra} ACIS-S synthetic spectrum (1 Ms of exposure time) of a plasma in CIE with ejecta abundance set to 3 for all the elements except for Si, which is set to 300 and an electron temperature of 0.8 keV, displayed with black crosses in Fig. \ref{lineonly}. The Si abundance is so high that the corresponding emission lines dominate the whole spectrum. Even if this scenario is not realistic, because we have not included the contribution of the ISM X-ray emission yet, it allows us to notice that the corresponding spectrum does not show any spectral signatures related to the RRC emission, which is expected to be present on the basis of the study on the fluxes discussed in the previous section. This is because the poor resolution of the CCD spectrometer makes the He-Si and H-Si lines heavily broadened. The line broadening blurs the recombination edges, thus hiding the spectral signature of pure-metal ejecta emission. In fact, by ignoring only the Si line emission (Si contributions to the FF and FB continuum are still included), the resulting spectrum (in red in Fig. \ref{lineonly}) shows, as expected, a prominent edge of recombination at the characteristic energy of Si-RRC. 

\begin{figure}[!ht]
    \centering
    \includegraphics[scale=0.22]{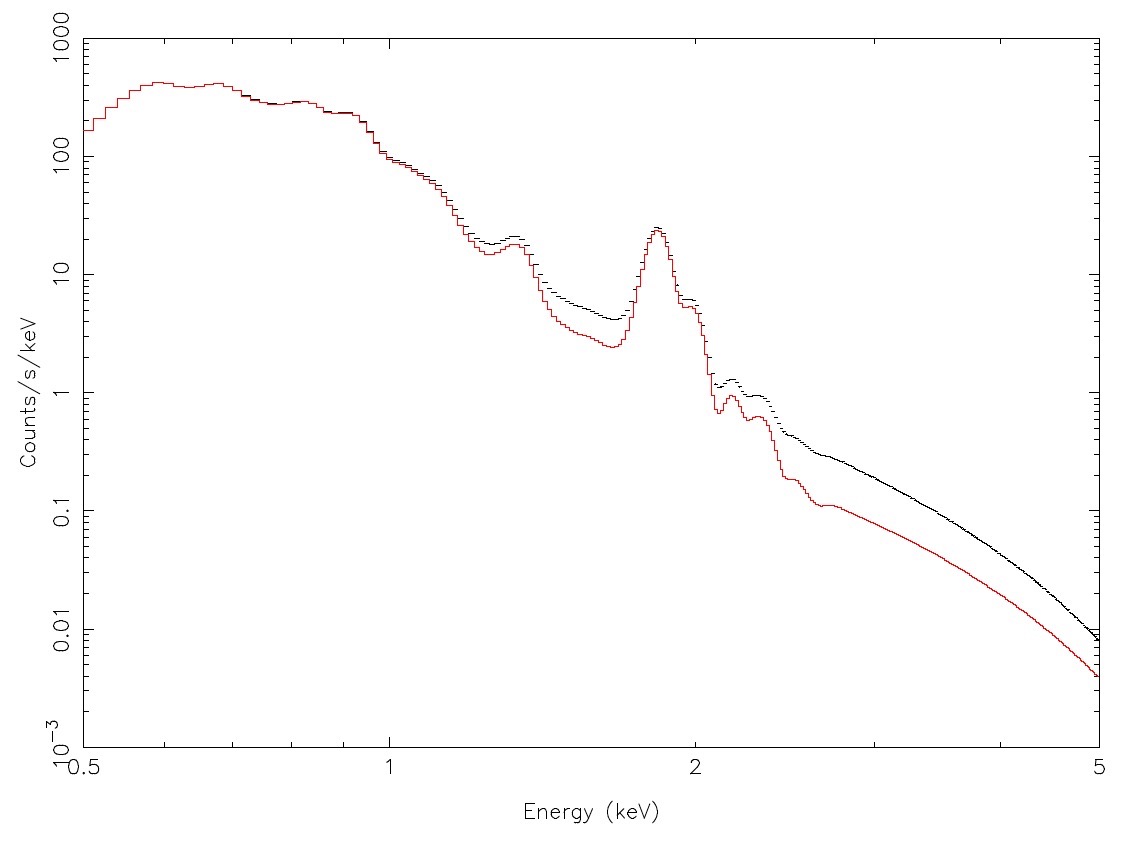}
    \caption{Synthetic ACIS-S spectra of ISM+pure-Si (pure-metal model, red) and ISM+Si-rich (mild-ejecta model, black) plasmas.}
    \label{SimMOS}
\end{figure}

The difficulty of observing enhanced FB emission and the contamination by shocked ISM hamper the identification of pure-metal ejecta in CCD spectra. As an example, we here show a synthetic SNR spectrum (a more realistic simulation, performed for a specific case, is presented in Sect \ref{introCasA}). We considered a spherical clump of Si-rich ejecta (Si abundance set to 300, as before) with radius $R_{clump}=0.5$ pc and temperature $kT_{clump}=0.8$ keV, surrounded by a colder ISM with temperature $kT_{ISM}=0.15$ keV. If we assume a density particle $n_e$ equal to 3 cm$^{-3}$ and pressure equilibrium between the clump and the ISM, and extract the spectrum from a box corresponding to a region of $8~\rm{pc} \times 8 ~\rm{pc}$ (in the plane of the sky) and extending $8~\rm{pc}$ along the line of sight, the ISM emission measure is 4 orders of magnitude larger than that of the clump (see Table \ref{bestsim} for details). This case is the one in which the effect is maximized but, as we will see also in Sect. \ref{introCasA}, this remains visible in the whole typical temperature range of SNRs and our conclusions do not depend on this choice. 
\begin{table}[!ht]
    \centering
    \begin{tabular}{c|c|c|c}
    \hline\hline
    Parameter & ISM & Pure-metal& Mild-ejecta\\
    \hline
    EM (cm$^{-3}$) & $1.6 \cdot 10^{59}$ & $1.5 \cdot 10^{55}$ & $1.5 \cdot 10^{57}$\\
    kT (keV)& 0.15 & 0.8& 0.8 \\
    Si Abundance& 1& 300& 3 \\ 
    \hline
    Si mass (M$_{\odot}$)&  /& 0.015 & 0.0016 \\
    Ejecta mass (M$_{\odot}$)& / & 0.06 & 0.6 \\
    \hline
    \end{tabular}
    \caption{Parameters of the ISM (CIE) plus pure-Si (CIE) or mild-Si models used for the spectral synthesis with the corresponding inferred Si and ejecta masses.}
    \label{bestsim}
\end{table}
We synthesized the ACIS-S X-ray spectrum using the CIE+CIE pure-metal model described in Table \ref{bestsim}. We assumed a distance of 1 kpc and an unrealistically high exposure time of 10$^8$ s in order to highlight the features. In Sect. \ref{introCasA} we will discuss cases with a more realistic exposure time. We adopted the response and ancillary files produced during the data reduction of \emph{Chandra} observation 114 (PI Holt). Figure \ref{SimMOS} shows, in red, the resulting synthetic ACIS-S spectrum.
As a comparison, we also produced a spectrum starting from the mild-ejecta model described in Table \ref{bestsim}, namely a model with the same parameters as those of the pure-metal model except only for the Si abundance, which is set to 3 (instead of 300), and for the ejecta EM, set to $1.6 \times 10^{57}$ cm$^{-3}$.
This new spectrum is shown, in black, in Fig. \ref{SimMOS}. The comparison between the two spectra does not reveal any clear difference that could be related to the pure-metal ejecta emission (even with an unrealistically high exposure time of 10$^8$ s). However, as we can see from the derived masses in Table \ref{bestsim}, in the pure-metal ejecta the total mass estimate goes down, but the mass of the considered element, Si in this case, goes up. We have two different models, with two different masses prediction, which are substantially indistinguishable. Spectra in Fig. \ref{SimMOS} confirm that the degeneracy between abundance and emission measure is a serious issue which is intrinsically due to the instrumental characteristics of the CCD cameras and does not depend on the statistics of the observation.
The ability to close off and identify the presence of FB contributions offer a unique diagnostic tool to assess whether the spectrum is coming from a highly enriched, but still hydrogen dominated plasma, or from a pure-metal ejecta plasma.

\subsection{Resolve}
Here, we furtherly show that the degeneracy between abundance and emission measure is intrinsically due to the instrumental characteristics of the CCD cameras and does not depend on the statistics of the observation. We repeated the spectral simulations discussed above by folding the pure-metal model and the mild-ejecta model with the Resolve response matrix\footnote{Resolve response and ancillary files used are xarm\_res\_h5ev\_20170818.rmf and xarm\_res\_flt\_fa\_20170818.arf, available on https://heasarc.gsfc.nasa.gov/docs/xrism/proposals/}. 

\begin{figure}[!ht]
    \centering 
    \includegraphics[scale=0.24]{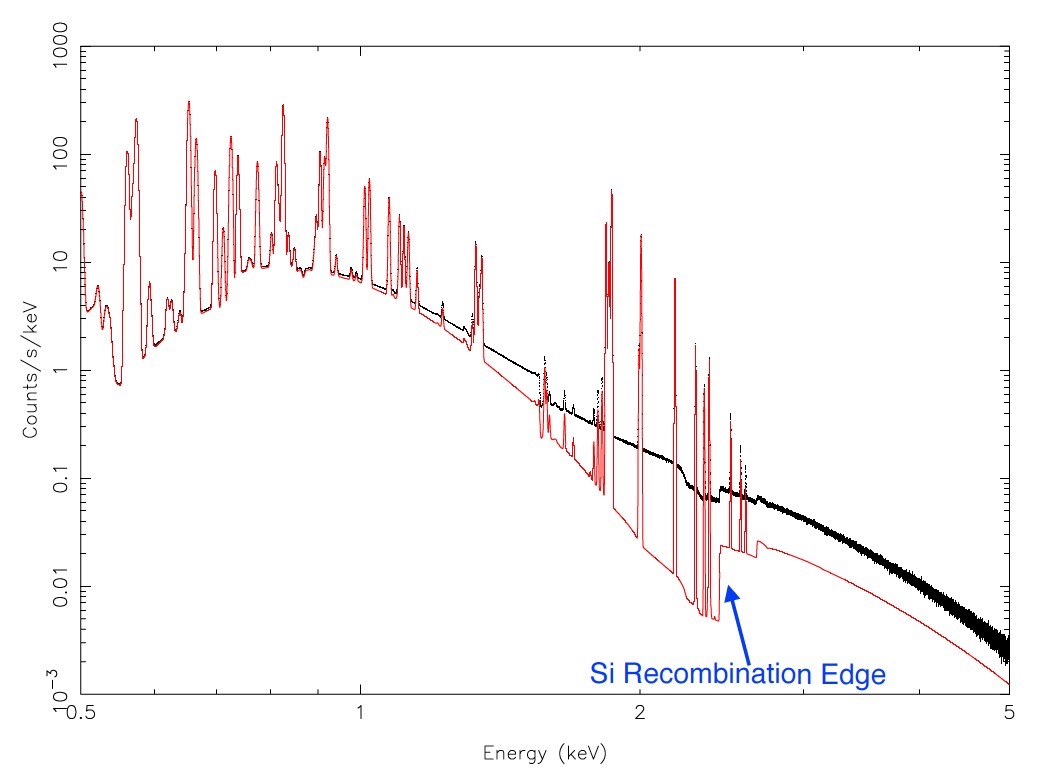}
    \caption{Same as Fig. \ref{SimMOS} but the model is folded through the Resolve response matrix.}
    \label{SimMicro}
\end{figure}

Figure \ref{SimMicro} shows the pure-metal case, in red, and the mild-ejecta case, in black assuming again an exposure time of 10$^8$ s. 
Thanks to the high spectral resolution of the microcalorimeter, it is now possible to observe a clear spectral difference between the two scenarios. As expected on the basis of the study presented in Sect. 2, a bright edge of recombination shows up at $\sim2.5$ keV (i.e., the He-Si RRC typical energy) when the abundance of Si is 300. Figure \ref{SimMicro} also clearly shows that the recombination edge and the RRC are much dimmer in the mild-ejecta case.
We stress that, in the pure-metal ejecta regime, even if the bremsstrahlung emission from the shocked ISM enhances the continuum emission and strongly reduces the equivalent width of emission lines, the RRC still emerges above the FF emission at energies $>2.5$ keV. Therefore, according to our simulations, the enhanced FB emission is a better tracer of pure-metal ejecta than the line equivalent width.

High resolution spectrometers like Resolve (and, in the future, X-IFU, on board the Advanced Telescope for High-Energy Astrophysics, ATHENA) are therefore capable to pinpoint the enhancement in the FB emission associated with a plasma with extremely high metallicity.

\section{Pure-metal ejecta in Cas A}
\label{introCasA}

 The simulations described above show that the enhancement in the RRC emission can be a strong signature of pure-metal ejecta and that such a spectral signature can be detected with high resolution spectrometers, while being almost impossible to be observed with CCD detectors. We here apply our diagnostic tool to a real case, by focusing on Cas A. In particular, we aim at understanding whether it will be possible to pinpoint the pure-metal ejecta emission with the Resolve spectrometer.
 
 Cas A is one of the brightest and most studied SNR. It is a young (330 years old, \citealt{tfv01}) SN IIb-type remnant (\citealt{kbu08})  at a distance of 3.4 kpc (\citealt{rhf95}) which shows many asymmetries and an overall clumpiness (\citealt{rhf95}, \citealt{hl03}, \citealt{vl03}, \citealt{hlb04}, \citealt{drs10} \citealt{hl12}, \citealt{mf13}, \citealt{lph14}). In particular, \cite{hl12} performed a detailed survey of the ejecta distribution in Cas A highlighting the presence of three large-scale Fe-rich clumps. They also confirmed the existence of an Fe-rich cloudlet, previously detected by \cite{hrb00} and \cite{hl03}, located within the southeastern clump (indicated by a red box in Fig. \ref{casadata}). In this cloudlet the relative Fe$/$Si abundance is $\sim20$, while Fe$/$Si$\sim5$ in the other Fe-rich regions of Cas A.

Here, we take advantage of the 3D hydrodynamic (HD) simulation of Cas A performed by \cite{omp16} (hereafter O16). This state-of-the-art simulation models the evolution of Cas A from the immediate aftermath of the supernova to the three-dimensional interaction of the remnant with the ambient environment. In particular, we adopted the model configuration that best describes the observed ejecta distribution (run CAS-15MS-1ETA in O16). This model reproduces the observed average expansion rate of the remnant and the shock velocities, and constraints the post-explosion anisotropies responsible for the observed structure and chemical distribution of ejecta. The model can reproduce very well the  shocked Fe distribution (both on large and relatively small spatial scales) while the Si (and S) mass seem to be slightly underestimated. We therefore focused on the Fe emission and adopted the HD simulation as a reference template to synthesize the expected X-ray emission. We point that the remnant evolution modeled by O16 clearly shows that large regions of Cas A are expected to be filled with pure Fe-rich ejecta.

We self-consistently produce synthetic \emph{Chandra/ACIS} and XRISM/Resolve spectra of the southeastern Fe-rich clump in Cas A from the 3D HD simulation. We compare the synthetic spectra with that observed by \emph{Chandra} and make predictions for the future Resolve observations.

\subsection{Data analysis}
\label{data}
\begin{figure}[!ht]
    \centering
    \includegraphics[scale=0.3]{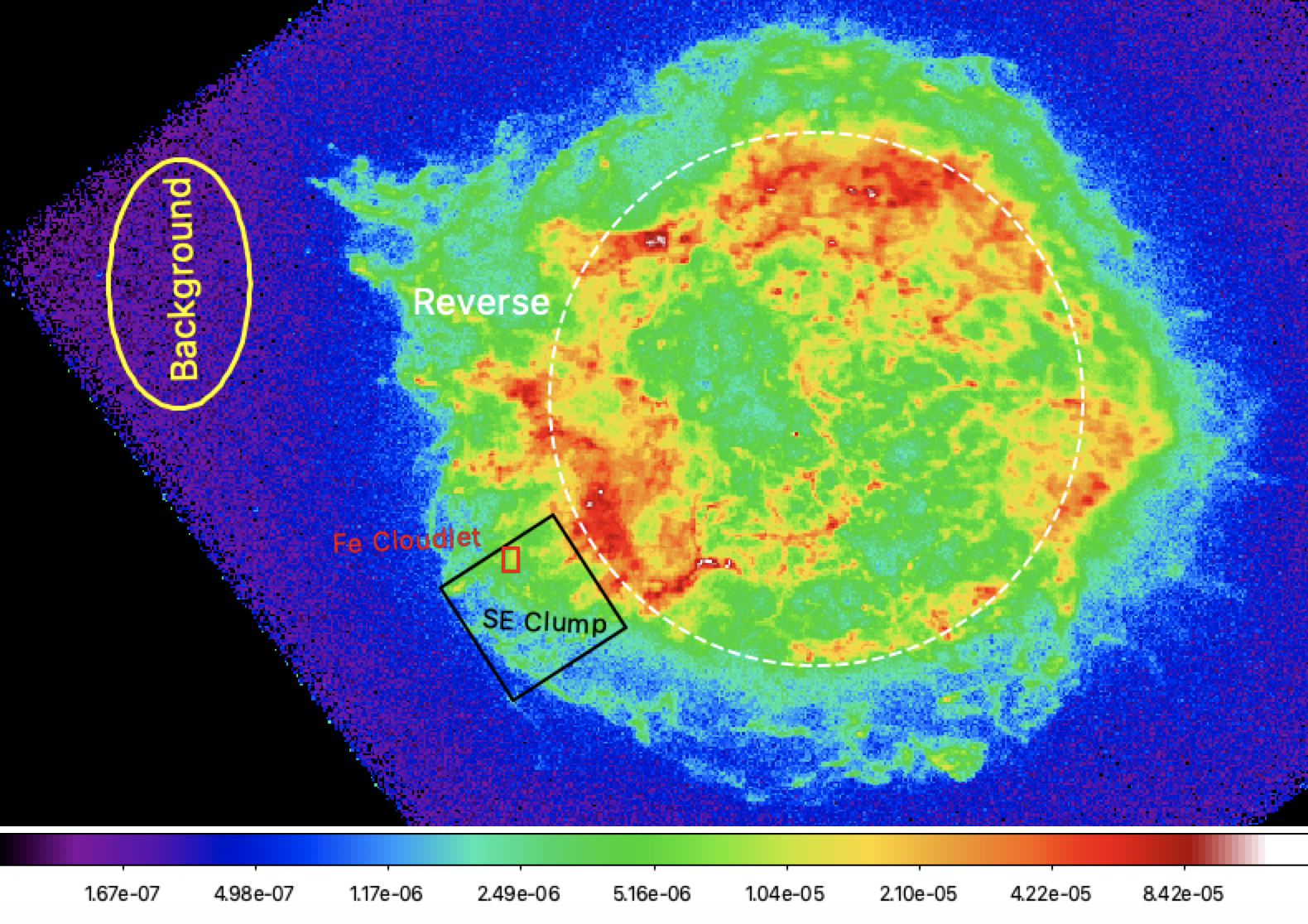}
    \caption{\emph{Chandra} count-rate image of Cas A in the $0.3-7$ keV energy band with a logarithmic color scale. The small red box marks the Fe-rich cloudlet, the black box indicates the region selected in the southeastern Fe-rich clump, the yellow ellipse shows the region chosen for the background and the dashed white circle marks the nominal position of the reverse shock.}
    \label{casadata}
\end{figure}

We analyzed the Chandra observation with ID 114 (PI Holt), performed on 30/01/2000.
We used the tool \emph{fluximage} to produce a count-rate image of Cas A with a bin size of $1''$. Fig. \ref{casadata} shows the count-rate image in the $0.3-7$ keV energy band together with the regions selected for the spectral extraction: the small red box marks the Fe-rich cloudlet, the black box indicates the region selected for the southeastern Fe-rich clump, and the yellow ellipse shows the region chosen for the background. 
Within the southeastern Fe-rich clump, nonthermal emission from the Cas A reverse shock has been detected (\citealt{gkr01}) and mapped accurately (\citealt{vh08}). Since synchrotron X-ray emission is not taken into account in our spectral synthesis code, we carefully selected the large box in the southeastern part of the shell by excluding the reverse shock.

By fitting the cloudlet spectrum we found the same best-fit parameters as \cite{hl03}, including the Fe/Si abundance ratio equal to $\approx 20$. In addition, we found that the absolute Fe abundance is not well constrained. In fact, two statistically equivalent fits having $\chi^2_{red}=1.55$ (136 d.o.f.) can be obtained with $A_{Fe} = 29$ (and EM$=1.4 \times 10^{55}$ cm$^{-3}$ for the ejecta component) and $A_{Fe} = 290$ (EM= $1.6 \times 10^{54}$ cm$^{-3}$). The cloudlet spectrum is shown, in red, in the upper panel of Fig. \ref{sinthspec}\footnote{Plots in Fig. \ref{sinthspec} are obtained through the spectral analysis software XSPEC (\citealt{arn96})}: the wide and bright structure at energies $\approx 1$ keV, reveals the presence of a remarkable complex of Fe L lines, but we do not see any Fe-RRC. As explained in Sect. \ref{allsynthesis}, the absence of this feature can be related to the instrumental characteristics of \emph{Chandra}. Upper panel of Fig. \ref{sinthspec} also shows, in black, the spectrum extracted from the black box of Fig. \ref{casadata} which presents a very strong Fe emission line complex, but with the addition of bright Si and S emission lines (the Fe/Si abundance ratio is lower than that in the Fe-rich cloudlet, being only $\sim 5$). This suggests the presence of both Fe-rich and Si-rich ejecta though we cannot exclude that the observed silicon emission may be somehow enhanced by the dust scattering in this region.  

\subsection{Self-consistent X-ray synthesis tool}

We developed a tool to self-consistently synthesize the thermal X-ray emission from the 3D HD simulation. In each cell of the computational domain we derive the local value of temperature, electron density, ionization parameter ($\tau$, defined as the time integral of the electron density computed from the impact with the shock front), total mass, and mass tracer of each element (see O16  for the list of isotopes included in the simulation). The mass tracer describes the mass of a given species as a fraction of the total mass in the cell. Our tool extracts all the aforementioned quantities for each cell and we use these values as input parameters for the non-equilibrium of ionization optically thin plasma model \emph{neij} (\citealt{kj93}) based on the atomic database SPEXACT 2.07.00, within SPEX. In particular, for each atomic species $i$, we derive the local value of ion and electron density and synthesize the corresponding ``pure-$i$" spectrum in each computational cell. The latter step is done by putting abundance of all elements except $i$ equal to 0, while $i$ abundance is set to 1. We then sum the resulting spectra over all the species by weighting each term for the corresponding emission measure. Finally, we sum the spectra of all the computational cells within a given region of the domain to derive the global spectrum of a selected area in the numerical simulation. All the spectra are filtered through the photoelectric absorption by the interstellar medium, with the appropriate column density (for this region $n_H=1.5 \times 10^{22}$ cm$^{-2}$, \citealt{hl12}) and are estimated assuming a distance of 3.4 kpc (\citealt{rhf95}). Each spectrum can be folded through the instrumental response matrix (i.e. \emph{Chandra}/ACIS-S or XRISM/Resolve). All the spectra are binned using the \emph{optimal bin} tool present in SPEX (\citealt{spex}).
By selecting the ACIS-S response matrix, we can directly compare the synthetic spectra with those observed, which is crucial for a fine-tuning of the selection of the region for the synthesis.

\subsection{Synthesis of Cas A spectra}
\label{synthdata}

\begin{figure}[!ht]
    \centering
    \includegraphics[scale=0.24]{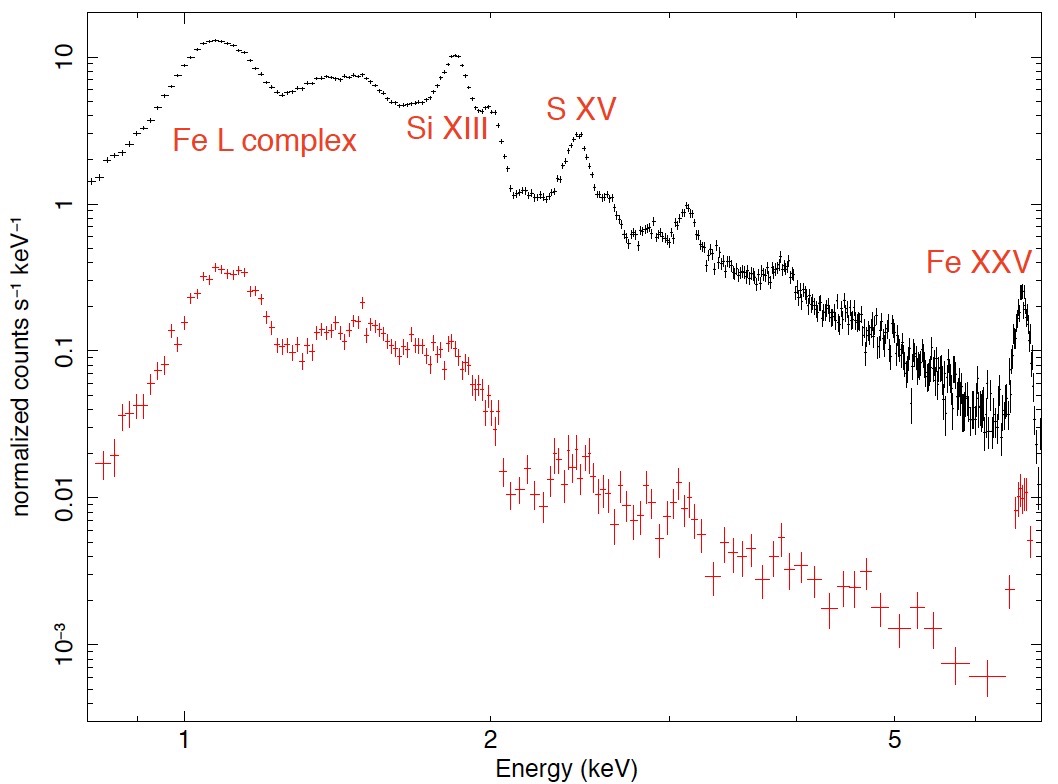}
    \includegraphics[scale=0.24]{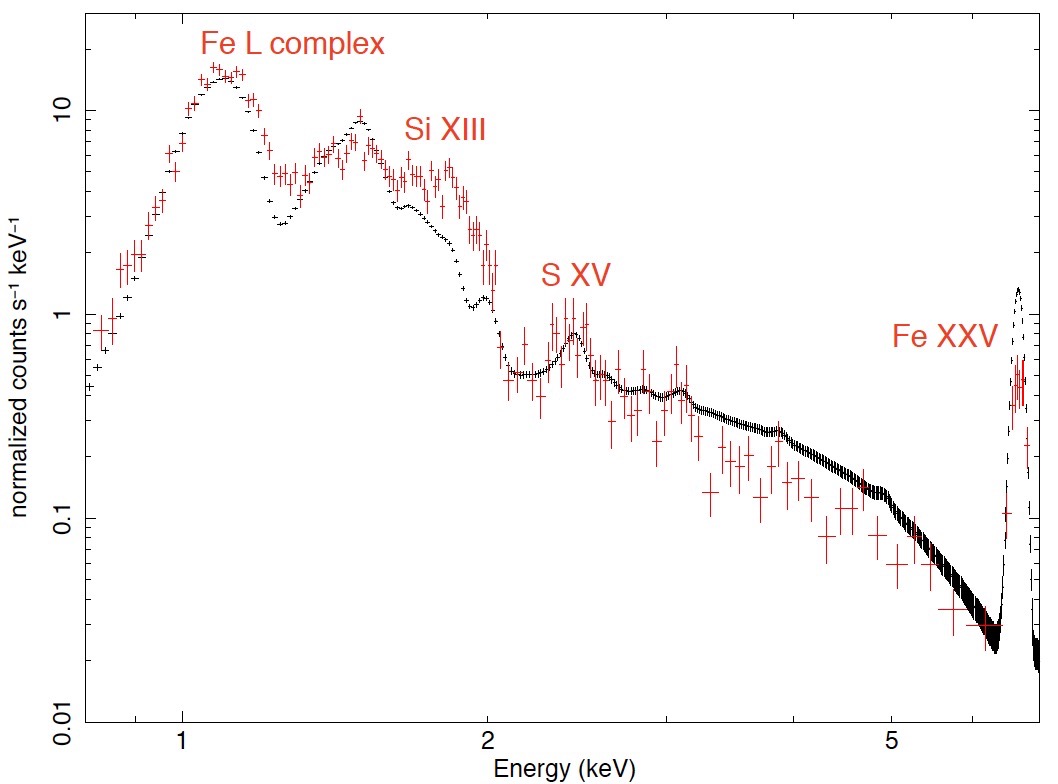}
    \caption{\emph{Upper panel:} \emph{Chandra} ACIS-S spectra extracted from the southeastern Fe-rich clump (black box in Fig. \ref{casadata}) and from the Fe-rich cloudlet (red box in Fig. \ref{casadata}) shown by black and red crosses, respectively. \emph{Lower panel:} \emph{Chandra} ACIS-S synthetic spectrum clump derived from the HD simulation of the southeastern Fe-rich (black crosses). Red crosses show the renormalized observed spectrum of the Fe-rich cloudlet shown in the upper panel for easy comparison with the synthetic spectrum.}
    \label{sinthspec}
\end{figure}

 We selected a region in the southeastern Fe-rich clump with the same size as the black box in Fig. 7, chosen for the actual \emph{Chandra} data (the spatial resolution of XRISM is not good enough to resolve the Fe-rich cloudlet marked in red in Fig. \ref{casadata}). The resulting synthetic \emph{Chandra} ACIS-S spectrum, obtained assuming an exposure time of 1 Ms, is shown, in black, in the lower panel of Fig. \ref{sinthspec}. The spectrum clearly shows signatures of bright Fe line emission, broadly similar to those actually observed in the corresponding region. At odds with the observations, the synthetic spectrum does not show very bright emission lines from Si and S ions, thus indicating that intermediate mass elements are somehow under-represented in the simulation, in this particular region. We point out that we are not aiming at finding a perfect agreement between synthetic and observed spectra, but we are interested in providing reliable and robust predictions on the X-ray emission from Fe-rich ejecta. We will show below that the lack of bright Si and S emission lines in the synthetic spectra does not affect our conclusions. We here notice that the synthetic spectrum is a good proxy of the real Fe-rich ejecta emission, given that it is extremely similar to the actual spectrum of the Fe-rich cloudlet (red box in Fig. \ref{casadata}). This is shown in Fig. \ref{sinthspec}, where the observed and renormalized spectra of the cloudlet can be compared with that derived from the HD simulation.
We only note some discrepancies in the brightness of the Fe XXV emission line, which is slightly overestimated in the synthetic spectrum. This is due to the fact that, in this region, the simulated temperature is slightly ($\sim30\%$) higher than that observed. However, this excess does not affect our conclusions since both actual and modeled temperatures lie in the range in which the Fe FB to FF ratio is maximized (see Appendix A). We only expect a slightly narrower edge of Fe-RRC in the actual spectra with respect to that derived from the simulation.

As in the actual data, we found that it is not possible to constrain the Fe abundance in the ACIS-S spectrum of this region. In fact, two statistically equivalent fits having $\chi^2_{red}=1.06$ (138 d.o.f.) can be obtained with $A_{Fe} = 45$ (and EM=1.3 $\times 10^{57}$ cm$^{-3}$ for the ejecta component) and $A_{Fe} = 1200$ (EM=6 $\times 10^{55}$ cm$^{-3}$). The corresponding Fe mass is therefore highly uncertain, spanning from $M_{Fe}=2$ $\times 10^{-4}$ M$_{\odot}$ to $M_{Fe}=4 \times 10^{-3}$ M$_{\odot}$.

\begin{figure}[!ht]
    \centering
    \includegraphics[scale=0.245]{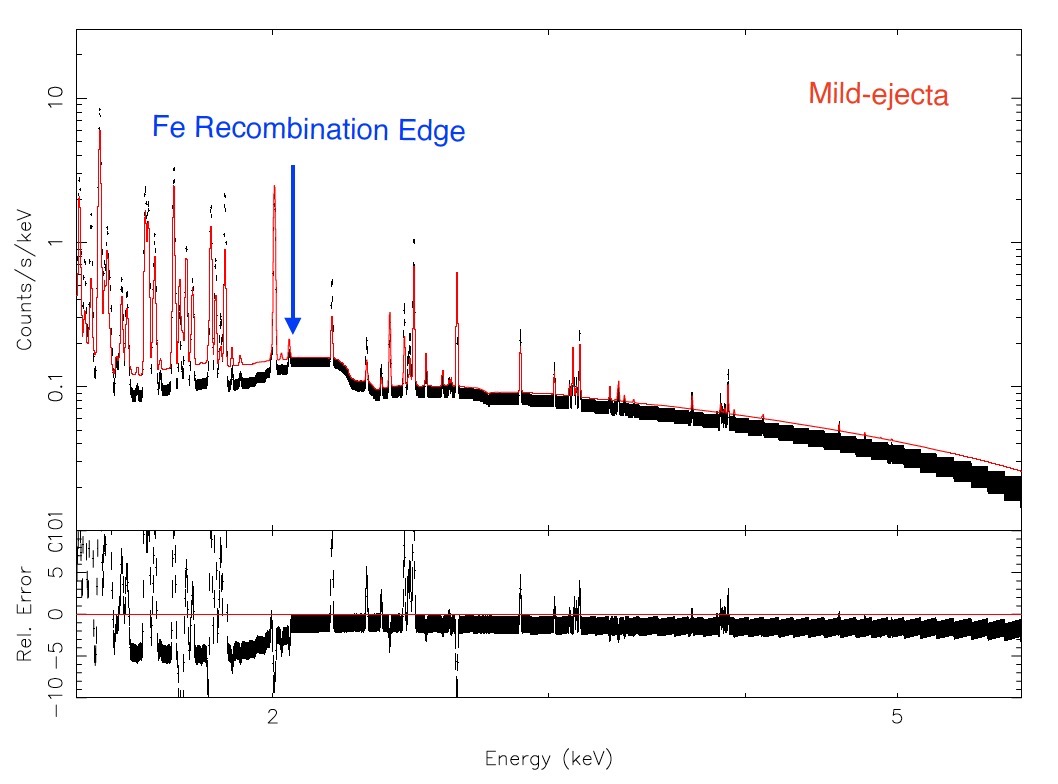}
    \includegraphics[scale=0.245]{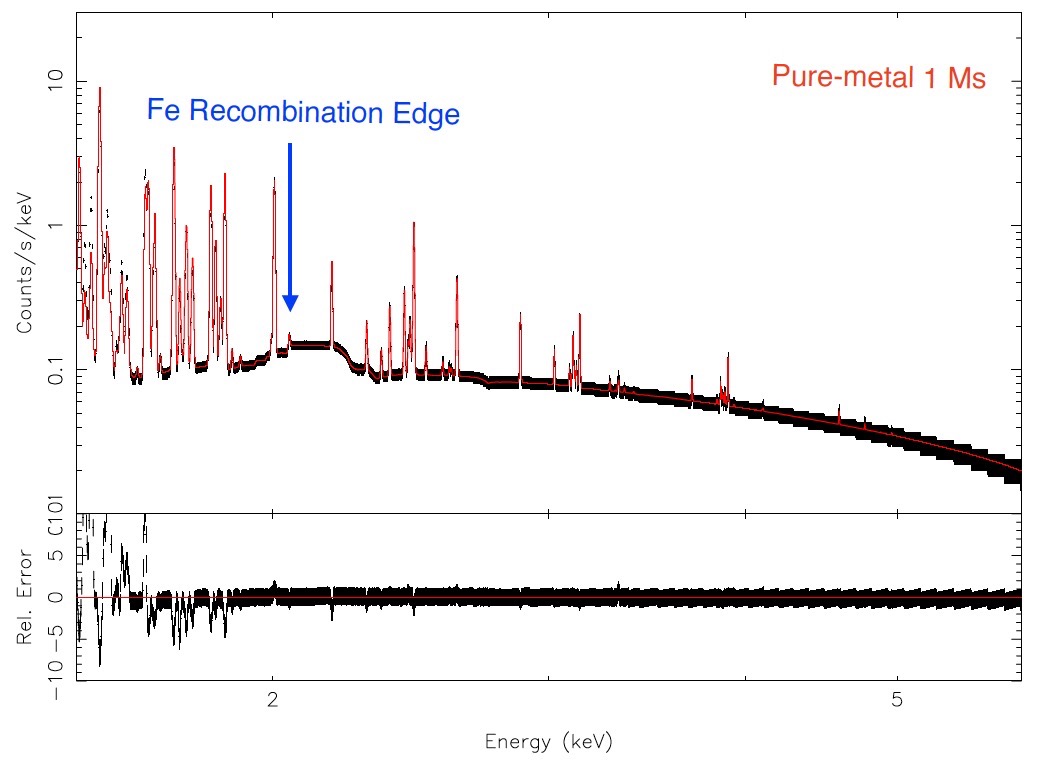}
    \includegraphics[scale=0.245]{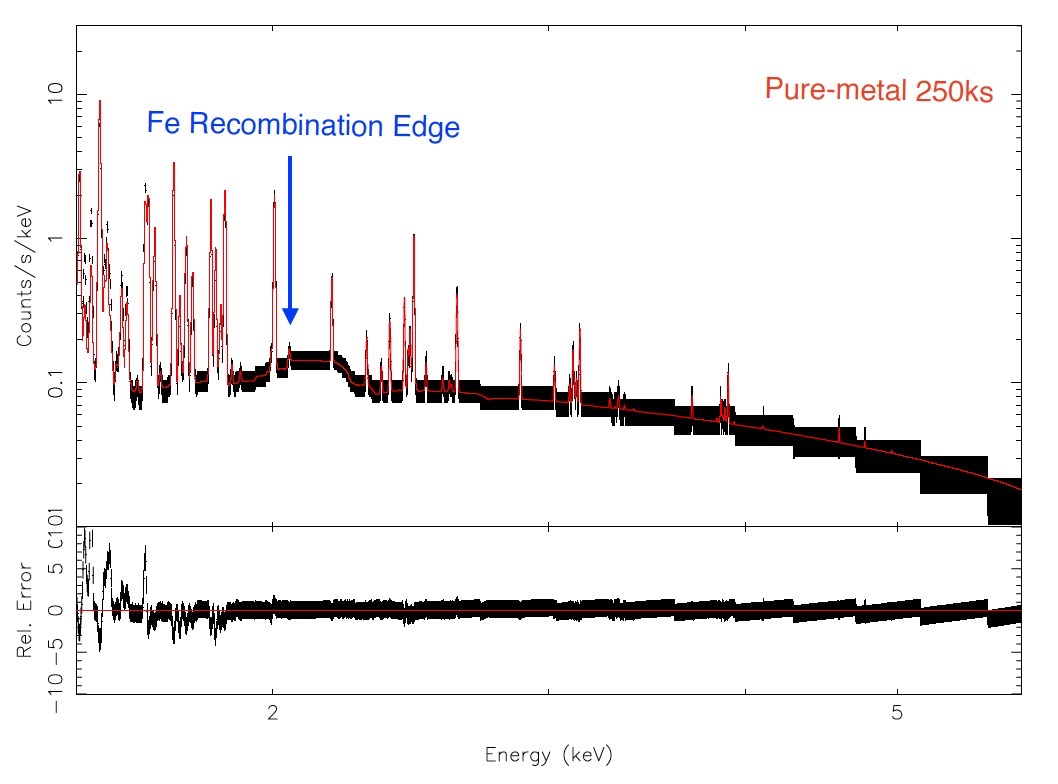}
    \caption{Synthetic XRISM spectra of the Fe-rich southeastern clump with the corresponding best fit models and residuals obtained for the mild-ejecta case (\emph{upper panel}), the pure-metal ejecta case (\emph{mid panel}) and the pure-metal ejecta case with an exposure time of 250 ks (\emph{lower panel}). See Tab. \ref{xrismspecfit} for details on the best fit parameters.}
    \label{synthxrism}
\end{figure}

We then synthesized the XRISM/Resolve spectrum from the same region, by assuming an exposure time of 1 Ms. The resulting spectrum is shown in
Fig. \ref{synthxrism}. The microcalorimeter spectrometer provides a superior spectral resolution, allowing us to clearly identify all the emission lines and spectral features. We here show that, by analyzing the Resolve synthetic spectrum, it is possible to unambiguously identify the enhanced RRC emission from the recombination of Fe ions, thus revealing the presence of pure-metal ejecta (as predicted in Sect. \ref{allsynthesis}).

 We fitted the spectrum with two isothermal components of an optically thin plasma in nonequilibrium of ionization associated with the shocked ISM and ejecta, respectively (hereafter NEI$+$NEI model). We considered two different scenarios: mild-ejecta and pure-metal ejecta. In the first case, we kept the Fe abundance fixed to 10 in the ejecta component (the spectrum with best fit model and residuals is shown in the upper panel of Fig. \ref{synthxrism}); in the second one we left the Fe abundance of the ejecta component free to vary (mid panel of Fig. \ref{synthxrism}). Best-fit parameters are shown in Table \ref{xrismspecfit}. 

\begin{table*}[!ht]
    \centering
    \begin{tabular}{c|c|c|c}
    \hline\hline
Parameter & NEI+NEI (mild-ejecta)& NEI+NEI (pure-metal ejecta)& NEI+NEI (pure-metal ejecta) 250ks \\
 \hline
 n$_H$ (10$^{22}$cm$^{-2}$) & \multicolumn{3}{c}{1.5(frozen)} \\
 \hline
 kT (keV)& 0.5$_{-0.001}^{+0.3}$ & 1.96$\pm$ 0.05& 1.62$_{-0.06}^{+0.11}$ , 
\\
 $\tau$ (10$^{11} $s/cm$^{3})$ &1.10$_{-0.05}^{+0.08}$ & 1.13$\pm$ 0.04 & 1.8$_{-0.2}^{+0.3}$\\ 

  EM (10$^{58}$ cm$^{-3}$)& 0.19$_{-0.17}^{+0.002}$ & 0.150$ \pm$ 0.004 & 0.20$_{-0.01}^{+0.02}$ \\
            \hline
  kT (keV)& 3.9$_{-0.15}^{+0.3}$  & 2.90$\pm$ 0.01& 2.89$\pm 0.03$\\ 
    $\tau$ ($10^{11}$ s/cm$^3)$ & 1.53$_{-0.08}^{+0.05}$& 2.51$\pm $ 0.02 & 2.51$_{-0.03}^{+0.05}$\\ 
    EM (10$^{58}$ cm$^{-3}$)& 0.34$_{-0.001}^{+0.04}$& 0.056$\pm$ 0.005 & 0.002\\
   Si  &0.415$_{-0.011}^{+0.003}$ & 0.77$\pm$ 0.08& 1 $\pm 1$\\
    S & 0.377$_{-0.013}^{+0.005}$  & 0.93$_{-0.07}^{+0.08}$& 2.0$_{-0.2}^{+2}$\\
    Ar &0.25 $\pm$ 0.02  & 0.4$\pm 0.2 $& 1$_{-1}^{+15}$\\
    Ca & 0.15$_{-0.03}^{+0.02}$& 0.49 $\pm$ 0.2& 1$_{-1}^{+7}$\\  
    Fe & 10 (frozen) &  119$_{-9}^{+11}$ &300$^{+400}_{-100}$\\
    \hline
    $\chi^2_{red}$ (d.o.f.) &11.85 (5005) & 2.11 (5004) &  0.65 (4709) \\
    \hline
    Counts & \multicolumn{2}{c}{1.4$\cdot 10^6$} & \\
    \hline
    \end{tabular}
    \caption{Best fit values for the fitting of the synthetic XRISM/Resolve spectrum of the Cas A southeastern region by adopting two thermal components, in the mild-ejecta and pure-metal ejecta scenarios. The abundances of the first component (associated with the ISM) are all frozen to 1. The rightmost column displays the best fit values obtained from the 250 ks spectrum.}
    \label{xrismspecfit}
\end{table*}

The spectral residuals and $\chi^2$ show that an Fe abundance $>100$  is necessary to properly fit the Fe RRC. The mild-ejecta scenario leads to strong residuals at the recombination edge energy and to a global misrepresentation of the spectrum. As already mentioned in Sect. \ref{chandrasynth}, we expect that in the pure-metal ejecta scenario the total mass estimate goes down, but the mass of the considered element, Fe in this case, goes up. Indeed, by fitting the spectrum synthesized with XRISM, we obtained $M_{Fe}=4.8_{-0.4}^{+0.5} \times 10^{-3}$ M$_{\odot}$, in good agreement with the value used in the simulation, 4.6 $\times 10^{-3}$ M$_{\odot}$. The high spectral resolution provided by the microcalorimeters allows us to reveal the enhanced RRC emission of pure metal ejecta and this clearly removes the degeneracy between abundance and emission measure in the fitting procedure leading to a correct estimate of the absolute Fe mass in this area. 

As explained above, the HD simulation provides an accurate description of the actual distribution of Fe-rich ejecta in Cas A and is able to reproduce the spectral features observed with \emph{Chandra}. However, the moderate spectral resolution of \emph{Chandra} (and CCD detectors, in general) does not allow us to confirm that pure-metal ejecta are indeed present in Cas A, as predicted by the O16 model. If pure-Fe ejecta are actually present in the southeastern limb of Cas A, we expect to pinpoint their presence with XRISM and to correctly derive their mass (see Sect. \ref{final}). 

As explained above, we expect that the actual spectrum will show more Si and S line emission than that predicted by our simulation (see Fig. \ref{sinthspec}). Therefore, we should observe brighter Si and S lines and a bright RRC from Si (at energies $E>2.5$ keV, see Fig. \ref{SimMicro}) and S ($E>3$ keV). These features will not mask out the Fe RRC, thus not affecting our conclusions. In fact, the Si and S lines will be well resolved by the XRISM spectrometer without contaminating the Fe RRC and the Si (and S) RRC; also, edges show up at energies higher than the Fe XXIV recombination energy (2.023 keV).

We note that the residuals at low energy in the lower panel of Fig. \ref{synthxrism} are due to our simplified spectral model that includes only one isothermal component for the ejecta. Indeed, our HD simulation shows a relatively broad distribution of plasma temperatures and ionization parameters in the region selected, but we are fitting the spectrum with only two temperatures and ionization parameters. A multi-component model provides a much better fit to the emission line complexes, but this is beyond the scope of this paper. In any case, even with this more complex spectral model, we verified that an Fe abundance $>100$ is always required to fit the spectrum.

So far, we showed synthetic spectra produced assuming an exposure time of 1 Ms. We are aware that it may be not possible to observe this region of Cas A for such a large amount of time. We then synthesized the same spectrum as in the upper panel of Fig. \ref{synthxrism} assuming an exposure time of 250 ks (lower panel in Fig. \ref{synthxrism}). The enhanced Fe-RRC remains visible in the spectrum and the pure-metal ejecta model describes the spectrum significantly better than the mild-ejecta model. An exposure time shorter than 250 ks may not be sufficient to unambiguously detect the pure-metal ejecta. 
\section{Discussion and conclusions}
\label{final}
In this paper we showed that spectral analysis carried with current X-ray detectors suffers from an intrinsic degeneracy in measuring the abundances using line dominated spectra from SNRs. We studied the behaviour of the X-ray emission processes at high abundances, identifying the enhanced RRC as the spectral signature of pure-metal ejecta. We showed that the low energy resolution of the current CCD detector hampers our understanding of the spectra, hiding the characteristic  RRC. We found that future microcalorimeter spectrometers, such as Resolve that will be onboard the XRISM telescope, can address this issue. We also presented a very promising case for the detection of pure-Fe emission in the southeastern clump of Cas A, by using both \emph{Chandra} data and 3D HD simulation to derive quantitative predictions on the expected Resolve spectrum. The resulting synthetic spectra show bright Fe-RRC at energy of around 2 keV and the fits performed on such spectra can definitely pinpoint the presence of pure-metal ejecta, if any. 

\subsection{Implications for SNRs with overionized ejecta components}

\begin{figure}[!ht]
    \centering
    \includegraphics[scale=0.22]{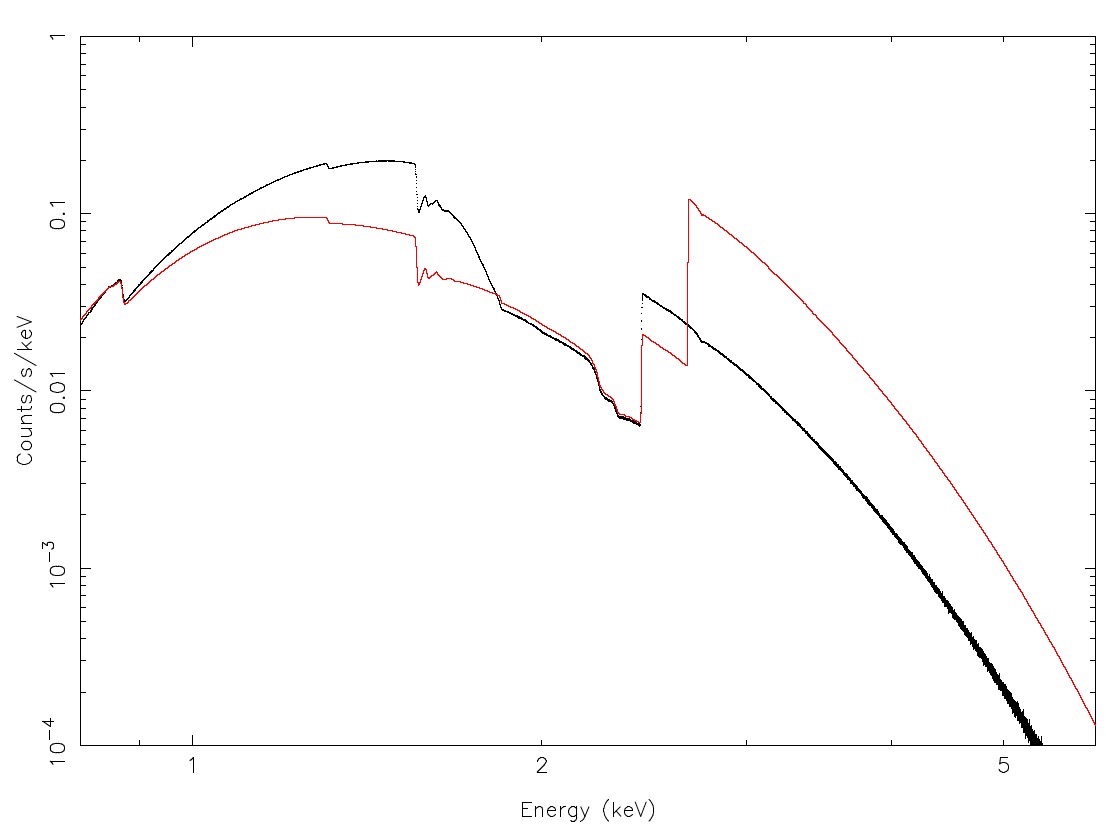}
    \includegraphics[scale=0.16]{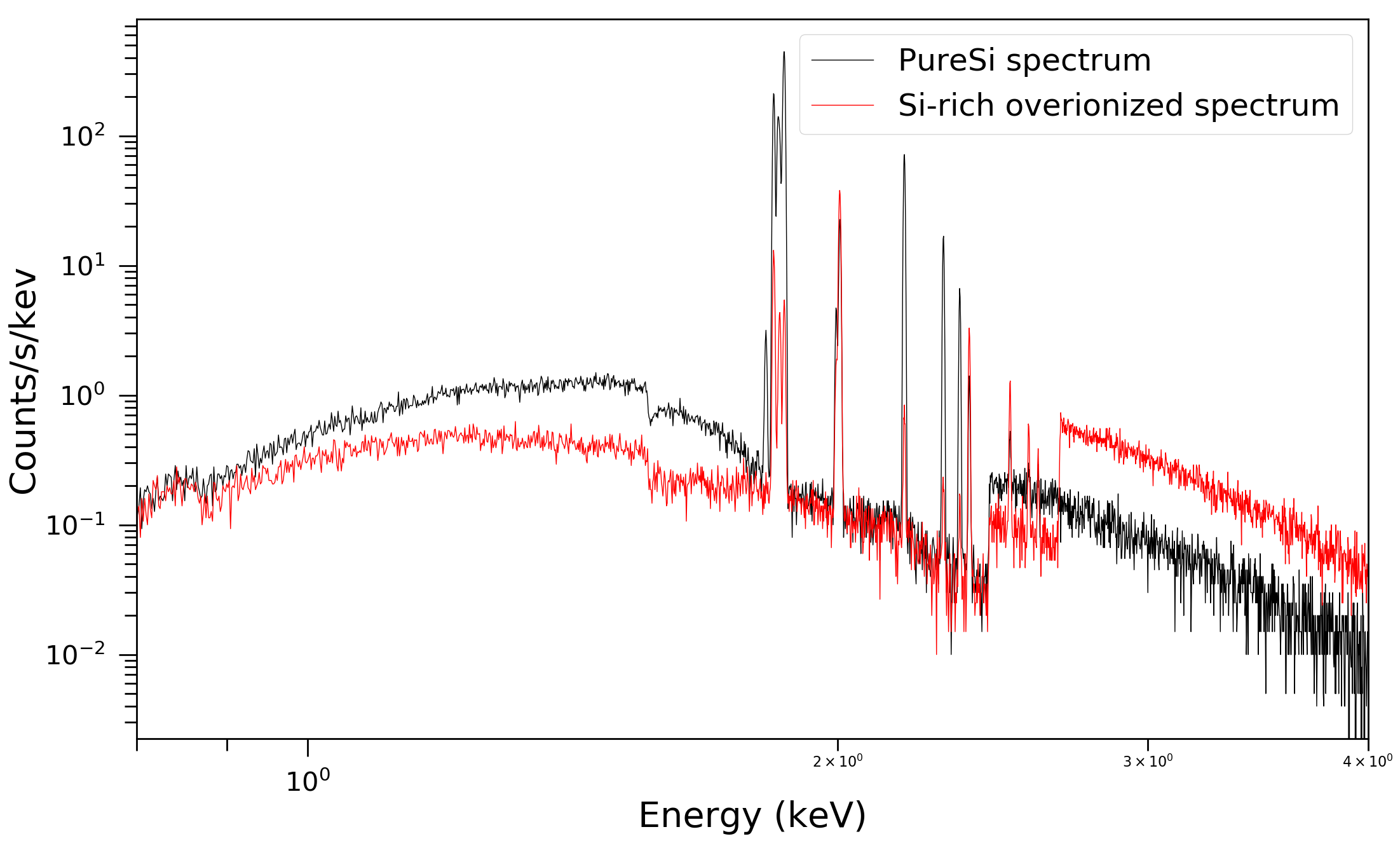}
    \caption{\emph{Upper panel}: in red, continuum-only spectrum of a typical overionized SNR (Si abundance set to 3, $\tau_{rec}=10^{11}$ s cm$^{-3}$, $kT_{\rm fin}$=0.5 keV and kT$_{0}$=5 keV). In black: continuum-only spectrum of a typical pure-ejecta plasma in CIE  (Si abundance set to 300 (black) and kT=0.5 keV). Both spectra are synthesized assuming an exposure time of 10$^8$ ks. \emph{Lower panel}: same as upper panel but assuming an exposure time of 100ks and including also the line emission.}
    \label{ovpure}
\end{figure}

The RRC features discussed here have a different physical origin with respect to those associated with overionized plasmas that have been detected so far in spectra of mixed-morphology SNR, namely those remnants which show bright radio emission in the outer shells and peaked X-ray emission in the internal ones. When a plasma in equilibrium of ionization undergoes a rapid cooling, the electron temperature is drastically lowered and the plasma degree of ionization  is higher than that expected in CIE for the observed electron temperature. In this scenario, the plasma is in a phase of strong recombination and the characteristic spectra show very bright recombination edges. The physical mechanism which causes the rapid cooling of the plasma is still debated and both thermal conduction with nearby molecular clouds (e.g. \citealt{otu20}) and adiabatic expansion (e.g. \citealt{zmb11}, \citealt{yok09}) have been proposed as viable cooling processes (even for the same object, see \citealt{mtu17} and \citealt{gmo18} for IC 443). We verified that the enhanced recombination features related to pure-metal ejecta can be distinguished from those associated with overionization. Upper panel of figure \ref{ovpure} shows, in red, the continuum emission of an overionized plasma with $\tau_{rec}=10^{11}$ s/cm$^3$, cooled down to $kT_{\rm fin}=$ 0.5 keV from an initial temperature $kT_0=5$ keV  (typical values for overionized plasmas in SNRs (\citealt{mtu17}, \citealt{gmo18}), and, in black, the continuum emission of ejecta in the pure-Si regime (Si abundance set to 300). The RRC produced by the overionized plasma is much stronger than that of pure-metal ejecta. This is because in the pure-metal ejecta regime the ratio FB/FF does not increase arbitrarily with the abundance, but reaches a saturation value (see Fig. \ref{RatioFBFF}), which is smaller than that obtained for the case of overionization. Moreover, in the case of overionization, the Si-edge is actually composed by two different edges: the one at lower energy is related to the He-like Si, while the second edge, which is the brightest, is associated with H-like Si. This is a characteristic feature of overionized plasmas, since the high degree of ionization leads to an increase in the H-like ions at expense of the He-like ions. Therefore, both the shape and the intensity of the RRC and edges make it possible to discriminate between pure-metal ejecta and overionized plasmas.

Nevertheless, the case of a pure-metal ejecta overionized plasma needs also to be considered. This scenario may be realistic for the Galactic SNR W49B (\citealt{oky09}, \citealt{mbd10}, \citealt{zmb11}, \citealt{lpr13}, \citealt{zv18}). The combination of the two effects (pure-metal plasma and overionization) further increases the contribution of the RRC emission and needs to be taken into account to correctly estimate the ionization parameter. The line emission has a crucial role in this scenario. In the lower panel of Fig. \ref{ovpure}, we show the same synthetic XRISM spectra displayed in the upper panel of Fig. \ref{ovpure} assuming an exposure time of 100ks and including also line emission. Even with this realistic exposure time we can easily distinguish the different shapes of the RRCs and discriminate between pure-metal ejecta and overionized plasma. Moreover, the line emission is completely different in the two spectra. In the recombining scenario the Si H-like lines dominate the spectrum and the He-like ones are significantly fainter. At the same time, the overall brightness of Si lines is higher in the pure-metal scenario, as we expect because of the high abundance. In conclusion, a careful analysis of the line emission is needed to correctly distinguish between the pure-metal ejecta and the recombining scenarios.

\subsection{Recovering the absolute mass thanks to RRC with Resolve spectra}
The capability of detecting pure-metal ejecta emission is crucial to recover the correct mass of the atomic species in the ejecta. We here show that from the synthetic XRISM spectrum of Cas A (discussed in the previous section) it is possible to properly derive the Fe mass. We compared the Fe mass calculated from the best-fit models of the synthetic XRISM and \emph{CHANDRA} spectra with the correct value, known from the HD simulation. In order to estimate the filling factor of ejecta and ISM from the analysis of the syntetic spectra, we followed the recipe by \cite{bms99} under the assumption of pressure equilibrium. As explained in Sect. \ref{synthdata}, by fitting the \emph{Chandra} synthetic spectrum, we found that the Fe abundance can vary from a few tens to more than $10^3$. By fitting the spectrum synthesized with XRISM, we instead obtained  $M_{Fe}=4.8_{-0.4}^{+0.5} \times 10^{-3}$ M$_{\odot}$. From the simulation, we know that the correct value of the Fe mass in the region is 4.6 $\times 10^{-3}$ M$_{\odot}$. Despite of the lower statistics, by fitting the 250 ks XRISM spectrum we obtain $M_{Fe}=6_{-2}^{+5} \times 10^{-3}$ which is consistent with the correct value. We stress that an exposure time shorter than 250 ks could lead to a non unambiguous detection of the pure-metal ejecta feature. By analyzing the XRISM spectra, we remarkably recover the real mass of the Fe-rich ejecta without the degeneracy present in the \emph{Chandra} analysis. The analysis of CCD spectra can be misleading in the estimation of Fe mass up to a factor of 20. This comparison further confirms that in order to correctly evaluate the ejecta mass it is crucial to accurately study the RRC of the elements and high-resolution spectrometers are needed. This is critical since the element mass is used in the comparison with the theoretical nucleosynthesis yields.

\bibliographystyle{aa}
\bibliography{references}

\begin{thebibliography}{46}
\expandafter\ifx\csname natexlab\endcsname\relax\def\natexlab#1{#1}\fi

\bibitem[{{Arnaud}(1996)}]{arn96}
{Arnaud}, K.~A. 1996, in ASP Conf. Ser. 101: Astronomical Data Analysis
  Software and Systems V, 17

\bibitem[{{Bocchino} {et~al.}(1999){Bocchino}, {Maggio}, \&
  {Sciortino}}]{bms99}
{Bocchino}, F., {Maggio}, A., \& {Sciortino}, S. 1999, \aap, 342, 839

\bibitem[{{DeLaney} {et~al.}(2010){DeLaney}, {Rudnick}, {Stage}, {Smith},
  {Isensee}, {Rho}, {Allen}, {Gomez}, {Kozasa}, {Reach}, {Davis}, \&
  {Houck}}]{drs10}
{DeLaney}, T., {Rudnick}, L., {Stage}, M.~D., {et~al.} 2010, \apj, 725, 2038

\bibitem[{{Frank} {et~al.}(2015){Frank}, {Burrows}, \& {Park}}]{fbp15}
{Frank}, K.~A., {Burrows}, D.~N., \& {Park}, S. 2015, \apj, 810, 113

\bibitem[{{Gotthelf} {et~al.}(2001){Gotthelf}, {Koralesky}, {Rudnick}, {Jones},
  {Hwang}, \& {Petre}}]{gkr01}
{Gotthelf}, E.~V., {Koralesky}, B., {Rudnick}, L., {et~al.} 2001, \apjl, 552,
  L39

\bibitem[{{Greco} {et~al.}(2018){Greco}, {Miceli}, {Orlando}, {Peres}, {Troja},
  \& {Bocchino}}]{gmo18}
{Greco}, E., {Miceli}, M., {Orlando}, S., {et~al.} 2018, \aap, 615, A157

\bibitem[{{Helder} \& {Vink}(2008)}]{vh08}
{Helder}, E.~A. \& {Vink}, J. 2008, \apj, 686, 1094

\bibitem[{{Hughes} {et~al.}(2000){Hughes}, {Rakowski}, {Burrows}, \&
  {Slane}}]{hrb00}
{Hughes}, J.~P., {Rakowski}, C.~E., {Burrows}, D.~N., \& {Slane}, P.~O. 2000,
  \apjl, 528, L109

\bibitem[{{Hwang} \& {Laming}(2003)}]{hl03}
{Hwang}, U. \& {Laming}, J.~M. 2003, \apj, 597, 362

\bibitem[{{Hwang} \& {Laming}(2012)}]{hl12}
{Hwang}, U. \& {Laming}, J.~M. 2012, \apj, 746, 130

\bibitem[{{Hwang} {et~al.}(2004){Hwang}, {Laming}, {Badenes}, {Berendse},
  {Blondin}, {Cioffi}, {DeLaney}, {Dewey}, {Fesen}, {Flanagan}, {Fryer},
  {Ghavamian}, {Hughes}, {Morse}, {Plucinsky}, {Petre}, {Pohl}, {Rudnick},
  {Sankrit}, {Slane}, {Smith}, {Vink}, \& {Warren}}]{hlb04}
{Hwang}, U., {Laming}, J.~M., {Badenes}, C., {et~al.} 2004, \apjl, 615, L117

\bibitem[{{Kaastra} \& {Jansen}(1993)}]{kj93}
{Kaastra}, J.~S. \& {Jansen}, F.~A. 1993, \aaps, 97, 873

\bibitem[{{Kaastra} {et~al.}(1996){Kaastra}, {Mewe}, \&
  {Nieuwenhuijzen}}]{spex}
{Kaastra}, J.~S., {Mewe}, R., \& {Nieuwenhuijzen}, H. 1996, in UV and X-ray
  Spectroscopy of Astrophysical and Laboratory Plasmas, ed. K.~{Yamashita} \&
  T.~{Watanabe}, 411--414

\bibitem[{{Krause} {et~al.}(2008){Krause}, {Birkmann}, {Usuda}, {Hattori},
  {Goto}, {Rieke}, \& {Misselt}}]{kbu08}
{Krause}, O., {Birkmann}, S.~M., {Usuda}, T., {et~al.} 2008, Science, 320, 1195

\bibitem[{{Kumar} {et~al.}(2012){Kumar}, {Safi-Harb}, \& {Gonzalez}}]{ksg12}
{Kumar}, H.~S., {Safi-Harb}, S., \& {Gonzalez}, M.~E. 2012, \apj, 754, 96

\bibitem[{{Lee} {et~al.}(2014){Lee}, {Park}, {Hughes}, \& {Slane}}]{lph14}
{Lee}, J.-J., {Park}, S., {Hughes}, J.~P., \& {Slane}, P.~O. 2014, \apj, 789, 7

\bibitem[{Lide(2003)}]{lid03}
Lide, D. 2003, CRC Handbook of Chemistry and Physics, 84th Edition, CRC
  HANDBOOK OF CHEMISTRY AND PHYSICS (Taylor \& Francis)

\bibitem[{{Liedahl}(1999)}]{lie1999}
{Liedahl}, D.~A. 1999, {The X-Ray Spectral Properties of Photoionized Plasma
  and Transient Plasmas}, ed. J.~{van Paradijs} \& J.~A.~M. {Bleeker}, Vol. 520
  (Lecture Notes in Physics), 189

\bibitem[{{Lodders} {et~al.}(2009){Lodders}, {Palme}, \& {Gail}}]{lpg09}
{Lodders}, K., {Palme}, H., \& {Gail}, H.~P. 2009, Landolt Boumlrnstein, 4B,
  712

\bibitem[{{Lopez} {et~al.}(2014){Lopez}, {Castro}, {Slane}, {Ramirez-Ruiz}, \&
  {Badenes}}]{lcs14}
{Lopez}, L.~A., {Castro}, D., {Slane}, P.~O., {Ramirez-Ruiz}, E., \& {Badenes},
  C. 2014, \apj, 788, 5

\bibitem[{{Lopez} {et~al.}(2013){Lopez}, {Pearson}, {Ramirez-Ruiz}, {Castro},
  {Yamaguchi}, {Slane}, \& {Smith}}]{lpr13}
{Lopez}, L.~A., {Pearson}, S., {Ramirez-Ruiz}, E., {et~al.} 2013, \apj, 777,
  145

\bibitem[{{Matsumura} {et~al.}(2017){Matsumura}, {Tanaka}, {Uchida}, {Okon}, \&
  {Tsuru}}]{mtu17}
{Matsumura}, H., {Tanaka}, T., {Uchida}, H., {Okon}, H., \& {Tsuru}, T.~G.
  2017, \apj, 851, 73

\bibitem[{{Mewe}(1999)}]{mew99}
{Mewe}, R. 1999, {Atomic Physics of Hot Plasmas}, ed. J.~{van Paradijs} \&
  J.~A.~M. {Bleeker}, Vol. 520 (Lecture Notes in Physics), 109

\bibitem[{{Miceli} {et~al.}(2010){Miceli}, {Bocchino}, {Decourchelle},
  {Ballet}, \& {Reale}}]{mbd10}
{Miceli}, M., {Bocchino}, F., {Decourchelle}, A., {Ballet}, J., \& {Reale}, F.
  2010, \aap, 514, L2+

\bibitem[{{Miceli} {et~al.}(2008){Miceli}, {Bocchino}, \& {Reale}}]{mbr08}
{Miceli}, M., {Bocchino}, F., \& {Reale}, F. 2008, \apj, 676, 1064

\bibitem[{{Miceli} {et~al.}(2006){Miceli}, {Decourchelle}, {Ballet},
  {Bocchino}, {Hughes}, {Hwang}, \& {Petre}}]{mdb06}
{Miceli}, M., {Decourchelle}, A., {Ballet}, J., {et~al.} 2006, \aap, 453, 567

\bibitem[{{Milisavljevic} \& {Fesen}(2013)}]{mf13}
{Milisavljevic}, D. \& {Fesen}, R.~A. 2013, \apj, 772, 134

\bibitem[{{Nakamura} {et~al.}(1999){Nakamura}, {Umeda}, {Nomoto}, {Thielemann},
  \& {Burrows}}]{num99}
{Nakamura}, T., {Umeda}, H., {Nomoto}, K., {Thielemann}, F.-K., \& {Burrows},
  A. 1999, \apj, 517, 193

\bibitem[{{Nomoto} {et~al.}(1997){Nomoto}, {Hashimoto}, {Tsujimoto},
  {Thielemann}, {Kishimoto}, {Kubo}, \& {Nakasato}}]{nht97}
{Nomoto}, K., {Hashimoto}, M., {Tsujimoto}, T., {et~al.} 1997, Nuclear Physics
  A, 616, 79

\bibitem[{{Okon} {et~al.}(2020){Okon}, {Tanaka}, {Uchida}, {Yamaguchi},
  {Tsuru}, {Seta}, {Smith}, {Yoshiike}, {Orlando}, {Bocchino}, \&
  {Miceli}}]{otu20}
{Okon}, H., {Tanaka}, T., {Uchida}, H., {et~al.} 2020, \apj, 890, 62

\bibitem[{{Orlando} {et~al.}(2016){Orlando}, {Miceli}, {Pumo}, \&
  {Bocchino}}]{omp16}
{Orlando}, S., {Miceli}, M., {Pumo}, M.~L., \& {Bocchino}, F. 2016, \apj, 822,
  22

\bibitem[{{Ozawa} {et~al.}(2009){Ozawa}, {Koyama}, {Yamaguchi}, {Masai}, \&
  {Tamagawa}}]{oky09}
{Ozawa}, M., {Koyama}, K., {Yamaguchi}, H., {Masai}, K., \& {Tamagawa}, T.
  2009, \apjl, 706, L71

\bibitem[{{Reed} {et~al.}(1995){Reed}, {Hester}, {Fabian}, \&
  {Winkler}}]{rhf95}
{Reed}, J.~E., {Hester}, J.~J., {Fabian}, A.~C., \& {Winkler}, P.~F. 1995,
  \apj, 440, 706

\bibitem[{{Sawada} \& {Koyama}(2012)}]{sk12}
{Sawada}, M. \& {Koyama}, K. 2012, \pasj, 64, 81

\bibitem[{{Sukhbold} {et~al.}(2016){Sukhbold}, {Ertl}, {Woosley}, {Brown}, \&
  {Janka}}]{sew16}
{Sukhbold}, T., {Ertl}, T., {Woosley}, S.~E., {Brown}, J.~M., \& {Janka}, H.-T.
  2016, \apj, 821, 38

\bibitem[{{Thielemann} {et~al.}(1996){Thielemann}, {Nomoto}, \&
  {Hashimoto}}]{tnh96}
{Thielemann}, F.-K., {Nomoto}, K., \& {Hashimoto}, M.-A. 1996, \apj, 460, 408

\bibitem[{{Thorstensen} {et~al.}(2001){Thorstensen}, {Fesen}, \& {van den
  Bergh}}]{tfv01}
{Thorstensen}, J.~R., {Fesen}, R.~A., \& {van den Bergh}, S. 2001, \aj, 122,
  297

\bibitem[{{Uchida} {et~al.}(2012){Uchida}, {Koyama}, {Yamaguchi}, {Sawada},
  {Ohnishi}, {Tsuru}, {Tanaka}, {Yoshiike}, \& {Fukui}}]{uky12}
{Uchida}, H., {Koyama}, K., {Yamaguchi}, H., {et~al.} 2012, \pasj, 64, 141

\bibitem[{{Vink} {et~al.}(1996){Vink}, {Kaastra}, \& {Bleeker}}]{vkb96}
{Vink}, J., {Kaastra}, J.~S., \& {Bleeker}, J. A.~M. 1996, \aap, 307, L41

\bibitem[{{Vink} \& {Laming}(2003)}]{vl03}
{Vink}, J. \& {Laming}, J.~M. 2003, \apj, 584, 758

\bibitem[{{Willingale} {et~al.}(2002){Willingale}, {Bleeker}, {van der Heyden},
  {Kaastra}, \& {Vink}}]{wbv02}
{Willingale}, R., {Bleeker}, J.~A.~M., {van der Heyden}, K.~J., {Kaastra},
  J.~S., \& {Vink}, J. 2002, \aap, 381, 1039

\bibitem[{{Woosley} \& {Weaver}(1995)}]{ww95}
{Woosley}, S.~E. \& {Weaver}, T.~A. 1995, \apjs, 101, 181

\bibitem[{{Yamaguchi} {et~al.}(2009){Yamaguchi}, {Ozawa}, {Koyama}, {Masai},
  {Hiraga}, {Ozaki}, \& {Yonetoku}}]{yok09}
{Yamaguchi}, H., {Ozawa}, M., {Koyama}, K., {et~al.} 2009, \apjl, 705, L6

\bibitem[{{Zhou} \& {Vink}(2018)}]{zv18}
{Zhou}, P. \& {Vink}, J. 2018, \aap, 615, A150

\bibitem[{{Zhou} {et~al.}(2019){Zhou}, {Vink}, {Safi-Harb}, \&
  {Miceli}}]{zvs19}
{Zhou}, P., {Vink}, J., {Safi-Harb}, S., \& {Miceli}, M. 2019, \aap, 629, A51

\bibitem[{{Zhou} {et~al.}(2011){Zhou}, {Miceli}, {Bocchino}, {Orlando}, \&
  {Chen}}]{zmb11}
{Zhou}, X., {Miceli}, M., {Bocchino}, F., {Orlando}, S., \& {Chen}, Y. 2011,
  \mnras, 415, 244

\end{thebibliography}

\appendix

\section{Spectral simulations for the Fe case}

\begin{figure}[!ht]
    \centering
    \includegraphics[scale=0.15]{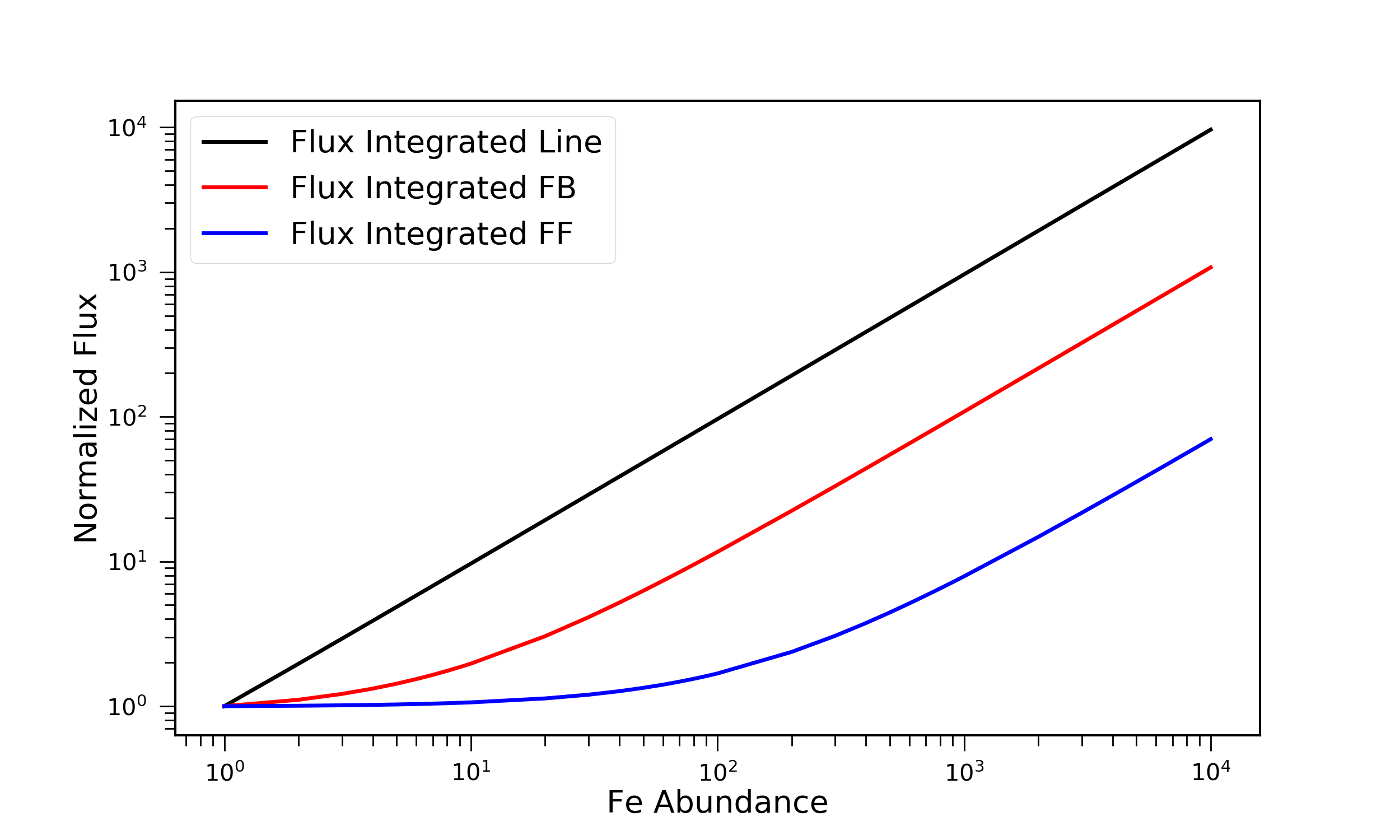}
    \includegraphics[scale=0.15]{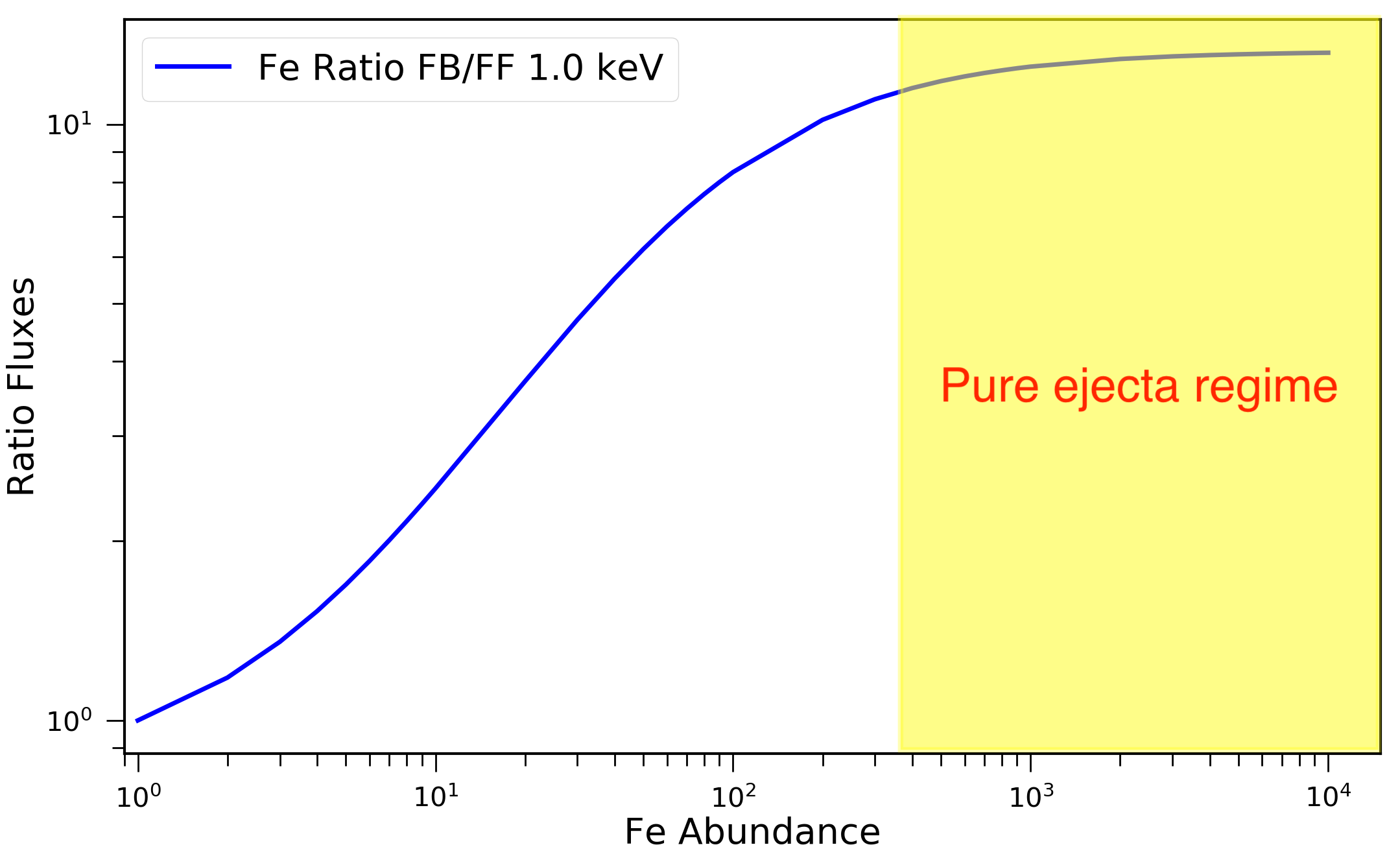}
    \caption{Upper panel: same as Fig. \ref{Flux_comparison} with Fe instead of Si. Lower panel: ratio between the FF and FB fluxes shown in the upper panel as a function of Fe abundance for a plasma temperature of 1 keV.}
    \label{fluxfe}
\end{figure}

\begin{figure}[!ht]
    \centering
    \includegraphics[scale=0.15]{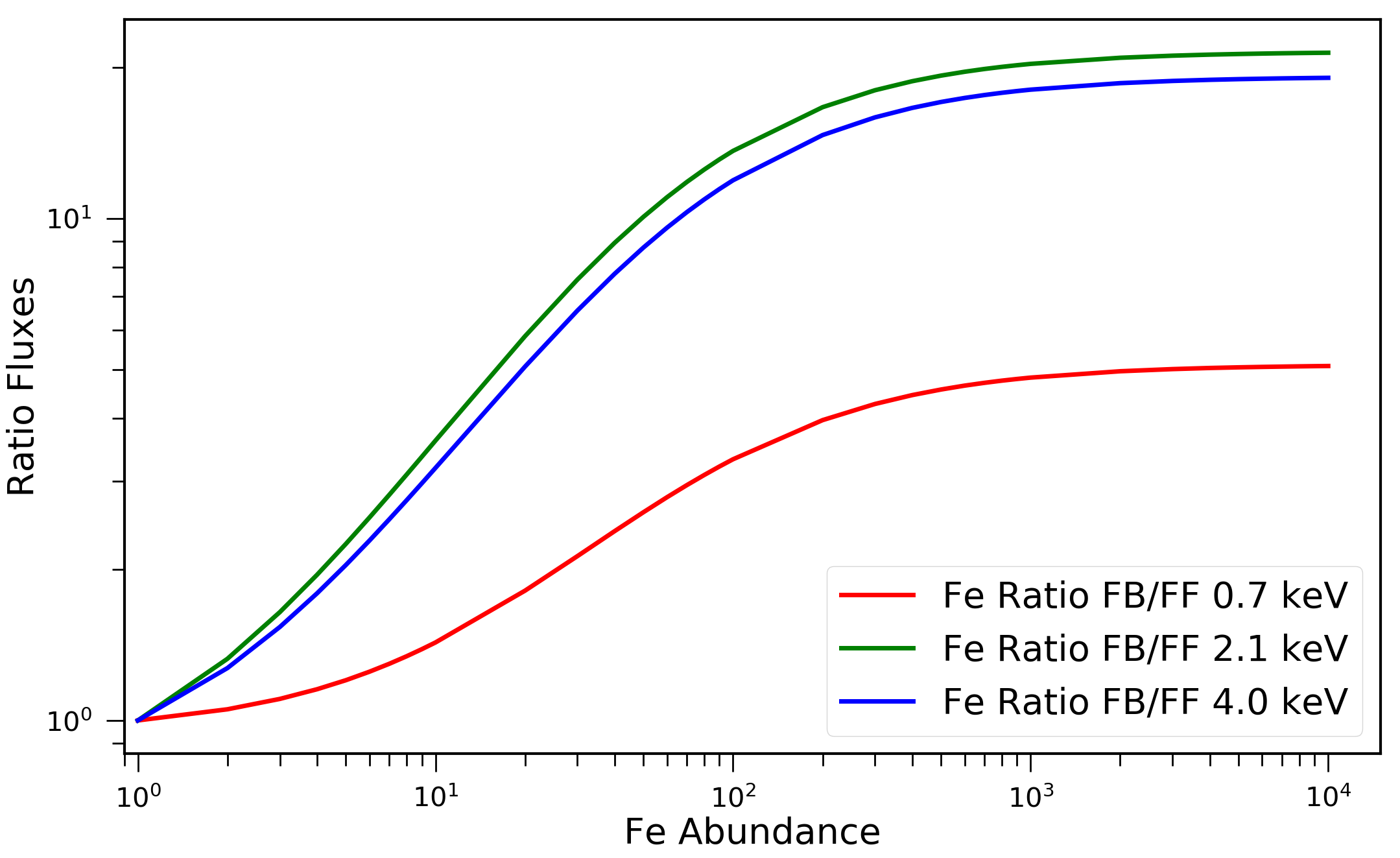}
    \caption{Same as lower panel of Fig. \ref{fluxfe} for plasma temperatures of 0.7 keV (red line), 2.1 keV (green) and 4.0 keV (blue).}
    \label{TdependFe}
\end{figure}

In this section we show results of the spectral simulations described in Sect. \ref{ss2}, by focusing on the Fe abundance. Fig. \ref{fluxfe} shows the integrated and normalized fluxes in the corresponding energy bands (see Table \ref{tab:bands}) of the various emission processes as functions of the Fe abundance, and the corresponding FB over FF ratio. The trend is the same as that already discussed for the Si case with the saturation occurring when the Fe abundance is close to one hundred. The study on the temperature dependence of the FB to FF ratio showed that the observability of the Fe-RRC is maximized when the electron temperature is in the range 1.5-3.5 keV (see Fig. \ref{TdependFe}). We also notice that there are different recombination edges, corresponding to the different Fe ionization states and that the brightest RRC depends on the temperature and on the degree of ionization of the plasma. In any case, the conclusions obtained for a specific Fe-RRC are valid also for the other ones.

\begin{figure}[!ht]
    \centering
    \includegraphics[scale=0.225]{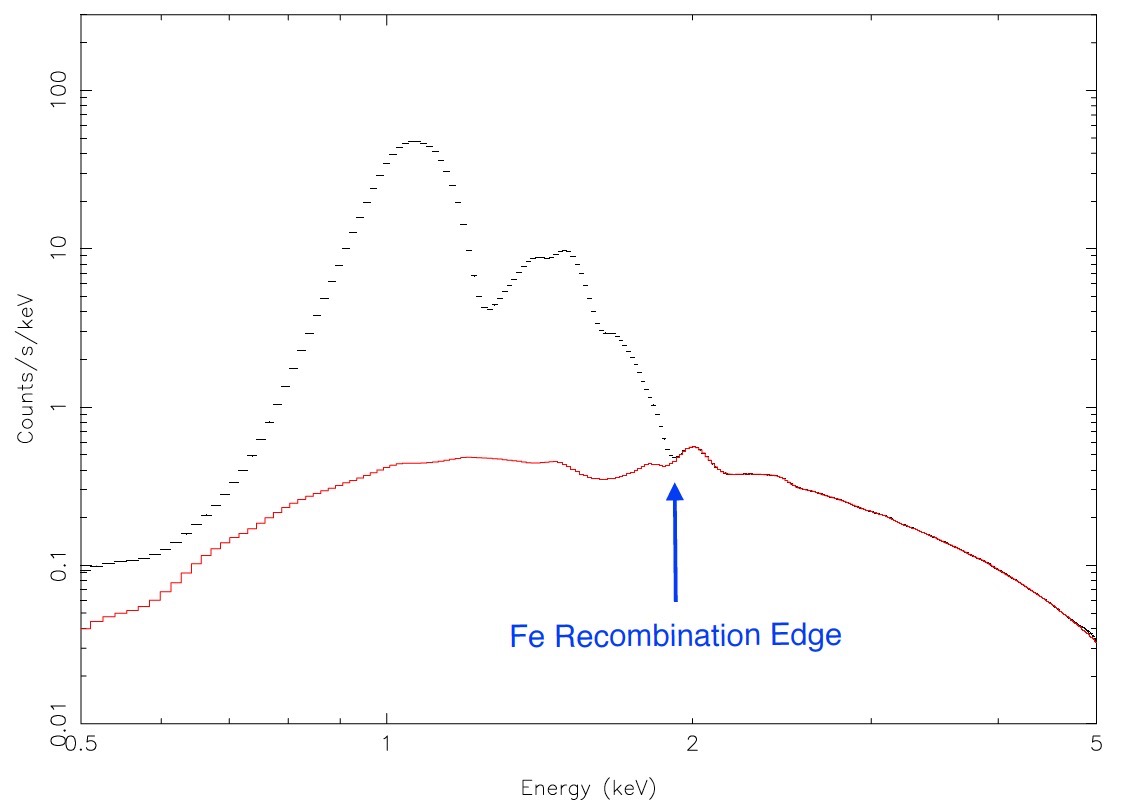}

    \caption{\emph{Black}: synthetic ACIS-S spectrum of a CIE plasma with abundance of all elements but Fe set to 3, Fe abundance set to 300, $kT=1.5$ keV and EM=$1.5\times10^{56}$ cm$^{-3}$. \emph{Red solid}: same spectrum but with the line emission subtracted.}
    \label{onlyfe}
\end{figure}

For all the syntheses we are going to discuss in this section we consider a distance of 1 kpc and an absorbing column density $n_H= 5 \times 10^{21} \rm{cm}^{-2}$, unless otherwise stated.  We synthesized a pure-Fe spectrum with abundances of all elements except Fe set to 3, Fe abundance set to 300, kT=1.5 keV and a particle density of 3 particles per cm$^3$. The resulting spectrum (black crosses in Fig. \ref{onlyfe}), synthesized assuming that the emission originates from a clump with radius $R_{clump}=$1.6 pc, showed the same issues faced with Si synthesis, due to the spectral resolution of the CCD detectors. We here chose a clump larger than that of the Si case because the considered RRC is in a part of the spectrum where the ISM emission is more significant (i.e., the ejecta FB emission is less visible). In any case, we verified that the goodness of our conclusions are not affected by this choice, as we also showed by discussing the realistic case of Cas A in the Sect. \ref{introCasA}. In addition, the scenario is even more complex because of the huge amount of Fe lines (from Fe XIV to Fe XXIV) present at energies around 1 keV. By ignoring the line emission of all the elements, we notice that the effective continuum contribution (red solid line of Fig. \ref{onlyfe}) to the emission shows an edge of recombination at the characteristic energy of Fe XXIV RRC. 

\begin{figure}[!ht]
    \centering
    \includegraphics[scale=0.24]{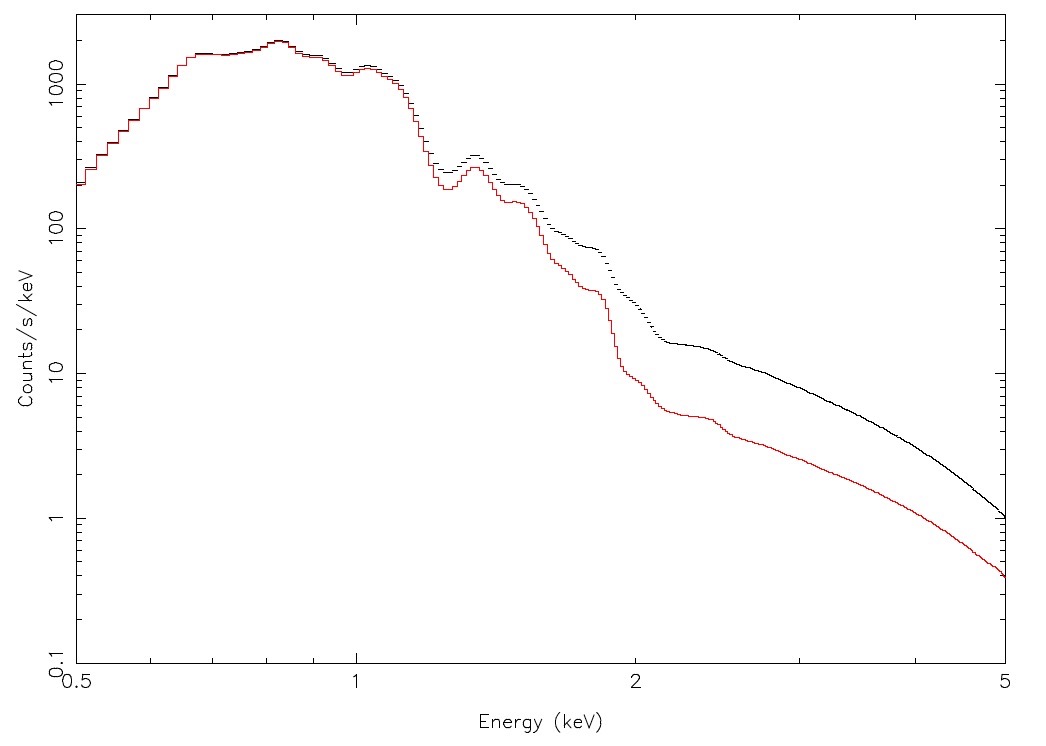}
    \includegraphics[scale=0.24]{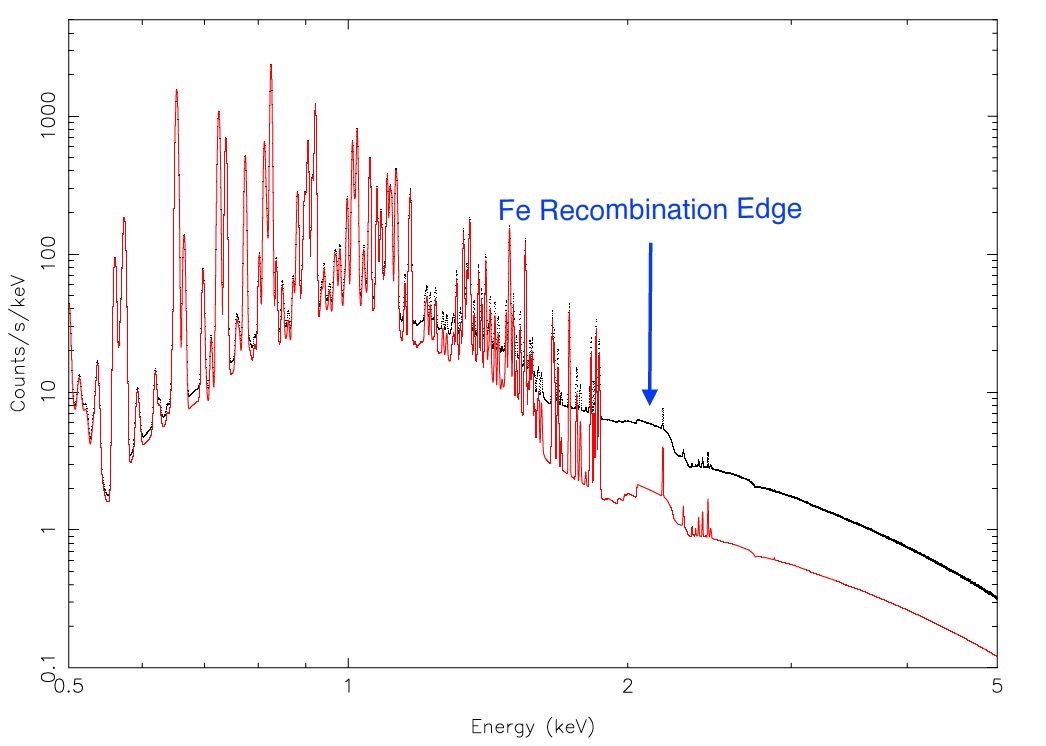}
    \caption{Upper panel: Synthetic ACIS-S spectra of ISM+pure-Fe (HighFe model, red) and ISM+Fe-rich (LowFe model, black) plasmas. Lower panel: same as upper panel but the model is folded with the Resolve response matrix.}
    \label{Fesim}
\end{figure}

We then produced synthetic spectra by adding another CIE component related to the ISM emission, considering the same region described in Sect. \ref{chandrasynth} for the Si case. We chose $T_{ISM}$= 0.23 keV and $T_{clump}$=1.5 keV. Even though the high temperature chosen for the clump maximizes the FB emission (as discussed above), we here show that CCD spectrometers cannot reveal the recombination edge. The parameters used for this pure-metal model and the derived Fe and total ejecta mass are summarized in Table \ref{bestsimfe}. The produced spectra, folded with ACIS-S/Chandra and Resolve/XRISM response matrices are shown, in red, in the upper and lower panels of Fig. \ref{Fesim}, respectively. 

\begin{table}[!ht]
    \centering
    \begin{tabular}{c|c|c|c}
    \hline\hline
    Parameter & ISM & Pure-metal & Mild-ejecta\\
    \hline
    Emission Measure (cm$^{-3}$) & $1.6 \cdot 10^{59}$ & $1.5 \cdot 10^{56}$ & $1.5 \cdot 10^{58}$  \\
    Temperature (keV)& 0.23 & 1.5 & 1.5 \\
    Fe Abundance& 1& 300& 3 \\ 
    \hline
    Fe mass (M$_{\odot}$) & / & 0.3& 0.04\\
    Ejecta mass (M$_{\odot}$)& / & 0.6& 6\\
    \end{tabular}
    \caption{Parameters of the ISM (CIE) plus pure-Fe (CIE) model used for the spectral synthesis. The abundance of other elements in the ejecta component is set to 0.}
    \label{bestsimfe}
\end{table}

We also produced a spectrum considering the mild-ejecta scenario (see again Table \ref{bestsimfe}) in which the Fe abundance is set to 3 (instead of 300) and the ejecta EM is set to $1.5 \times 10^{58} \, \mathrm{cm}^{-3}$. This spectrum is shown, in black, in both panels of Fig. \ref{Fesim}.

The results of the simulations performed on Fe are analogous to the Si ones and confirm that the instrumental broadening of the line completely hides the possible presence of RRC related to the pure-metal ejecta emission.

\end{document}